\documentclass[amssymb,prd,superscriptaddress,aps,nofootinbib,twocolumn,showpacs]{revtex4-1}
\usepackage{graphicx, epsfig, amssymb} 
\usepackage{amsmath, amsfonts}
\usepackage{bm} 
\usepackage[breaklinks]{hyperref}
\usepackage{color}
\usepackage{enumerate}
\usepackage{units}
\usepackage{multirow}

\def\be{\begin{equation}}
\def\ee{\end{equation}}
\def\beq{\begin{eqnarray}}
\def\eeq{\end{eqnarray}}

\begin{document}

\title{Evolution of a proto-neutron star with a nuclear many-body equation of
state: Neutrino luminosity and gravitational wave frequencies}


\author{Giovanni~Camelio}\email{giovanni.camelio@astro.su.se} \affiliation{Dipartimento di Fisica, ``Sapienza''
  Universit\`a di Roma \& Sezione INFN Roma1, P.A. Moro 5, 00185, Roma, Italy.}
  \affiliation{Astronomy and Oskar Klein Centre, Stockholm University, AlbaNova, SE-10691, Stockholm, Sweden.}

\author{Alessandro~Lovato}\affiliation{Physics Division, Argonne National Laboratory, Argonne, 
Illinois 60439, USA}

\author{Leonardo~Gualtieri}\email{leonardo.gualtieri@roma1.infn.it} \affiliation{Dipartimento di Fisica, ``Sapienza''
  Universit\`a di Roma \& Sezione INFN Roma1, P.A. Moro 5, 00185, Roma, Italy.}

\author{Omar~Benhar}\affiliation{Dipartimento di Fisica, ``Sapienza''
  Universit\`a di Roma \& Sezione INFN Roma1, P.A. Moro 5, 00185, Roma, Italy.}

\author{Jos\'e~A.~Pons}\email{jose.pons@ua.es} \affiliation{Departament de F\'isica Aplicada, Universitat d'Alacant,
  Apartat de Correus 99, 03080 Alacant, Spain.}

\author{Valeria~Ferrari}\email{valeria.ferrari@roma1.infn.it} \affiliation{Dipartimento di Fisica, ``Sapienza''
  Universit\`a di Roma \& Sezione INFN Roma1, P.A. Moro 5, 00185, Roma, Italy.}

\begin{abstract}
  In a core-collapse supernova, a huge amount of energy is released in the
  Kelvin-Helmholtz phase subsequent to the explosion, when the proto-neutron
  star cools and deleptonizes as it loses neutrinos. Most of this energy is
  emitted through neutrinos, but a fraction of it can be released through
  gravitational waves. We model the evolution of a proto-neutron star in the
  Kelvin-Helmholtz phase using a general relativistic numerical code, and
  a recently proposed finite temperature, many-body equation of state; from this we
  consistently compute the diffusion coefficients driving the evolution. To
  include the many-body equation of state, we develop a new fitting formula
  for the high density baryon free energy at finite temperature and
  intermediate proton fraction. We estimate the emitted neutrino signal,
  assessing its detectability by present terrestrial detectors, and we
  determine the frequencies and damping times of the quasi-normal modes 
  which would characterize the gravitational wave signal emitted in this stage.
\end{abstract}

\pacs{
}

\maketitle

\section{Introduction}\label{sec:intro}

When a star with mass greater than about $\unit[8]{M_\odot}$ exhausts its fuel,
the electron Fermi pressure can not prevent the collapse of the stellar core.
In a few milliseconds, the density of the collapsing core reaches the nucleon
density, the pressure due to the nucleon Fermi degeneracy and nuclear interaction
sets in, the collapse halts, and a shock wave is generated as the exterior core
layers bounce off the core.  Then, on a longer timescale the stellar core keeps
on contracting as it cools and deleptonizes, while the shock wave proceeds
through the stellar envelope.  This part of the evolution is known as the
Kelvin-Helmholtz phase, and the contracting stellar core is called proto-neutron
star (PNS). 
This phase lasts for tens of seconds, during which the PNS matter is opaque to neutrinos.
It has been shown that, after about 200~ms from the core bounce, the 
PNS evolution can be modeled as a
sequence of quasi-stationary configurations, where neutrino diffusion
determines the thermal and composition evolution of the hot
remnant~\citep{Burrows+Lattimer.1986, Keil+Janka.1995, Pons+.1999}.  When the
PNS has radiated about $\unit[10^{53}]{erg}\simeq \unit[0.1]{M_\odot}$ by
neutrinos, the temperature is low enough for the matter to become neutrino
transparent, and a neutron star is born.

The observation of a nearby supernova in the Large Magellanic Clouds in 1987, and the simultaneous detection of 19
neutrinos~\cite{Hirata+1987,Bionta+1987} have been milestones for both astrophysics and particle physics.  Since then,
impressive progresses have been made in the modeling of supernova (SN) explosions.  Numerical codes have been developed
to study the highly dynamical process of core collapse and core bounce. From the earlier 1D
simulations, multi-dimensionality has been extended to one+two and one+three, while including more and more complex
physical inputs (for a recent review, see e.g.~\cite{Janka:2016fox}). The effort in modeling the subsequent PNS phase has been
comparatively smaller, even though a considerable amount of energy is emitted in this phase.

Because of its much longer timescale, for many years, the complex core-collapse numerical codes have not been able to
describe the PNS phase.  Only recently core-collapse codes have been able to describe the PNS phase~\cite{Hudepohl+2010,
  Fischer+2010}, mainly with the aim of studying the nucleosynthesis processes due to the neutrino wind.

The quasi-stationary evolution of a PNS was firstly studied
in~\cite{Burrows+Lattimer.1986}.  After this first, seminal work in the past
years a number of papers have addressed several related issues, as for instance
the sensitivity of the PNS evolution and of the related neutrino signal to the
nuclear Equation of State (EoS)~\cite{Burrows.1988,
Pons+.1999,Keil+Janka.1995}, the possible delayed formation of black
holes~\cite{Burrows.1988,Pons+.1999}, convective effects in presence of
accretion~\cite{Burrows.1988,Roberts+2012, Takiwaki+Kotake+Suwa.2014,
Melson+Janka+Marek.2015, Muller.2015}, and nucleosynthesis due to the neutrino
wind~\cite{Roberts.2012}.

In addition, the frequencies at which a gravitational wave (GW) signal would be
emitted by an oscillating PNS have been computed
in~\cite{Ferrari+Miniutti+Pons.2003,Ferrari+.2004}, using quasi-equilibrium
configurations obtained from the evolutionary code of Pons {\it et
al.}~\cite{Pons+.1999}, based on a mean field EoS. In~\cite{Burgio+2011}, a
many-body EoS was employed, but the entropy and lepton fraction profiles were
included ``by hand'' in order to mimic a time evolution similar to that found
in~\cite{Pons+.1999}.
The entropy and lepton fraction profiles were included in a similar way
in~\cite{Sotani:2016uwn}, in order to mimic the the profiles obtained, in the
first second after bounce, by numerical core-collapse simulations. However,
the EoSs they employed (such as that of Lattimer and
Swesty~\cite{Lattimer+Swesty.1991}) are more appropriate to describe the
core-collapse phase than the PNS evolution. 
We remark that, up to now, finite temperature, many-body nuclear dynamics have
not been included in a consistent way (i.e., accounting for the modifications in
the neutrino cross sections) in PNS evolution.

In this paper we describe the results of  a new PNS evolutionary code and a
formula that allows to fit a general nucleonic EoS at finite temperature, as
the recently proposed many-body EoS of~\cite{Benhar+Lovato.2017,
Lovato+Benhar.prep}.  Using this code and three different EoSs (among which,
the many-body EoS proposed in~\cite{Benhar+Lovato.2017, Lovato+Benhar.prep}),
we study the PNS evolution during the Kelvin-Helmholtz phase. We estimate the
neutrino luminosity, and compute the frequencies and damping times of the PNS
quasi-normal modes (QNMs), which characterize the emitted GW signal.

The work is organized as follows. In Sec.~\ref{sec:EOS} we describe
the nucleonic EoSs adopted in this paper and a new nucleonic fitting
formula for the free energy.  In
Sec.~\ref{sec:diffusion} we describe how we compute the diffusion coefficient, 
and show how do we
effectively describe the baryon single-particle spectra by means of
effective masses and single-particle potentials.  In
Sec.~\ref{sec:evolution} we show the results of  our evolutionary
code, and discuss how the relevant quantities, which describe the
stellar structure and the neutrino luminosity, change in time. We also determine
the neutrino signal in the Super-Kamiokande III detector for our models.  In
Sec.~\ref{sec:GW} we describe the computation of the QNM frequencies and damping times, and we discuss how
the first QNMs change as the PNS evolve. We derive a relation between the frequencies of the fundamental mode and of
the first pressure mode, and the mean stellar density. In
Sec.~\ref{sec:conclusions} we draw our conclusions.  In
Appendix~\ref{app:fit} we provide the details of the fitting procedure
of the nucleonic EoS; in Appendix~\ref{app:code_checks} we discuss the
convergence of our PNS code and justify some of the approximations
made;  in Appendix~\ref{app:tables} we tabulate the frequencies and
damping times of the QNMs of the stellar configurations we consider.

Unless otherwise stated, we set to unity the speed of light, the Boltzmann constant, and
the gravitational constant $c=k_\mathrm{B}=G=1$. The ``microscopic'' masses,
like the bare and effective masses of neutron and proton, are given in MeV.
The ``macroscopic'' masses, that is, the PNS baryon and gravitational masses,
are given in terms of the Sun mass $M_\odot$.  We include the rest mass in the
chemical potential and in the energy density.

\section{The Equation of State}
\label{sec:EOS}
In this paper we compare three different finite-temperature nucleonic EoSs: a
mean field EoS, GM3~\cite{Glendenning.1985,Glendenning+Moszkowski.1991}; a
nuclear many-body EoS, CBF-EI, obtained using the correlated basis function
theory~\cite{Benhar+Lovato.2017, Lovato+Benhar.prep}; and a model based on the extrapolation from the
measured nuclear properties, LS-bulk~\cite{Lattimer+Swesty.1991}. In all EoSs
the leptonic part consists of a Fermi gas of non-interacting electrons,
positrons and neutrinos of all flavours, where neutrinos are treated as massless
particles.  The baryonic part consists of an interacting Fermi gas of protons
and neutrons.
We neglect the Coulomb force between protons (which is screened by the
electrons), we assume charge-independent nuclear interactions, and the proton
and neutron bare masses are set equal, $m_p\equiv m_n$.  Since we are interested
in the evolution of a proto-neutron star, pasta phases or a solid crust are not
included in our model.  We have checked \emph{a posteriori} that this
approximation is justified, since the PNS temperature is always above the
critical temperature for the formation of alpha particles, with the exception of
the end of the cooling phase, when this approximation is no longer accurate in
the region near the stellar surface (see Appendix~\ref{ssec:NSE}).

In the GM3 EoS, baryons\textemdash described by quantum
fields\textemdash interact through the exchange of bosons (the
$\sigma$, $\omega$ and $\rho$ mesons). The resulting equations of
motion are solved in the mean field approximation, which amounts to
treating mesons as classical fields. The LS-bulk EoS, specifically
designed to be easily implemented in stellar collapse simulations, is
based on a dynamical model constrained by nuclear phenomenology, and correspond to the
bulk part of the \citet{Lattimer+Swesty.1991} EoS. The
CBF-EI EoS (that stands for ``Correlated Basis Functions -- Effective
Interaction'') has been obtained within non-relativistic many-body theory,
using a realistic nuclear Hamiltonian, which includes the Argonne $v_6^\prime$
and the Urbana IX nuclear potentials. The
formalism of correlated basis functions and the cluster expansion
technique have been used to devise an effective nucleon-nucleon
potential, which includes the effects of both two- and three-nucleon
forces, as well as nuclear correlations. This effective potential is
well behaved and allows to describe both cold and hot matter, at
arbitrary proton fraction at the Hartree-Fock level.

It is easy and fast to compute the GM3 and LS-bulk EoSs during the simulation.  Conversely, due to their heavy
computational cost, this procedure cannot be adopted for many-body EoSs (like CBF-EI).  Therefore, one should resort
either to an interpolation, or to a fit.  Since we are studying the evolution of a PNS, we would need thermodynamical
consistency and continuity of the second order derivatives of the free energy~\cite{Swesty.1996}. It is difficult to
interpolate a table in a thermodynamically consistent way, because in a PNS the EoS is characterized by three
independent variables (see below).  Therefore, to describe the baryon interaction we will find, and use, a fitting
formula.

\subsection{Thermodynamical relations}
%
\label{ssec:thermo}

The first law of thermodynamics can be written in terms of an infinitesimal
variation of $f$, the free energy per baryon, as
\begin{equation}
\label{eq:df}
\mathrm df=-s\mathrm d T + \frac{P}{n_\mathrm{B}^2}\mathrm d n_\mathrm{B} +
\sum_i \mu_i \mathrm dY_i,
\end{equation}
with
\begin{align}
\label{eq:P}
P={}&\left.n_{\mathrm B}^2\frac{\partial f}{\partial n_\mathrm{B}}\right|_{T,\{Y_i\}},\\
\label{eq:s}
s={}& \left. -\frac{\partial f}{\partial T} \right|_{n_\mathrm{B},\{Y_i\}},\\
\label{eq:mu1}
\mu_i={}& \left. \frac{\partial f}{\partial Y_i}
\right|_{T,n_\mathrm{B},\{Y_{j\neq i}\}},
\end{align}
where $s$ is the entropy per baryon, $P$ the pressure, $T$ the temperature,
$n_\mathrm{B}$ the baryon number density, $\mu_i$ and $Y_i$ are the chemical potential
of particle $i$ and its particle fraction (i.e., the number of particles $i$ per
baryon) respectively.  Note that $f\equiv e-Ts$, $e$ being the energy per
baryon. In the following we will also use the energy density $\epsilon\equiv
en_\mathrm{B}$.  We remark that 
we include the rest mass in the energy and in the free energy, and therefore the
chemical potentials \emph{include the rest mass}.

Since the number fractions $\{Y_i\}$ are not independent variables, one should
consider the equation
\begin{equation}
\label{eq:mu2}
\mu_i=\left. \frac{\partial ( n_\mathrm{B} f )}{\partial n_i}
\right|_{T,\{n_{j\neq i}\}},
\end{equation} rather than  Eq.~\eqref{eq:mu1}.
If only neutrons and protons are present, Eq.~\eqref{eq:mu2} gives
\begin{align}
\label{eq:mup}
\mu_p={}&
f_\mathrm{B}+\frac{P_\mathrm{B}}{n_\mathrm{B}}+\left.(1-Y_p)\frac{\partial
f_\mathrm{B}}{\partial Y_p}\right|_{T,n_\mathrm{B}},\\
\label{eq:mun}
\mu_n={}& f_\mathrm{B}+\frac{P_\mathrm{B}}{n_\mathrm{B}}-\left.Y_p\frac{\partial
f_\mathrm{B}}{\partial Y_p}\right|_{T,n_\mathrm{B}},
\end{align}
where the subscript B means that we are considering only the baryon part of the
EoS.

\subsection{Baryon free energy fitting formula}
\label{ssec:fit}

In this section we shall discuss a fitting formula for the interacting part of
the baryon free energy; we remark that all thermodynamical quantities can be
obtained in terms of partial derivatives of the free energy. We shall not
consider here the kinetic part, which is the standard fermionic free energy (see
Sec.~\ref{ssec:eos_num}), and the leptonic free energy, which will also be
discussed in Sec.~\ref{ssec:eos_num}. In the following, the superscripts $I$ and
$K$ refer to the interacting and kinetic parts of the thermodynamical
quantities, respectively: \begin{equation}
f_\mathrm{B}(Y_p,T,n_\mathrm{B})=f_\mathrm{B}^K(Y_p,T,n_\mathrm{B})+f^I_\mathrm{B}(Y_p,T,n_\mathrm{B}).
\label{eq:f} \end{equation} To begin with, we discuss the dependency of
$f^I_\mathrm{B}$ on the proton fraction $Y_p$.  In a zero-temperature EoS, the
baryon free energy coincides with the baryon energy $e_\mathrm{B}$. Its
dependence on $Y_p$ is well approximated~\cite{Bombaci+Lombardo.1991} by
\begin{multline}
\label{eq:e_I}
e_\mathrm{B}^I(Y_p,T=0) = e^I_\mathrm{SNM}+(1-2Y_p)^2(
e^I_\mathrm{PNM}-e^I_\mathrm{SNM})\\
=4Y_p(1-Y_p)e^I_\mathrm{SNM} + (1-2Y_p)^2e^I_\mathrm{PNM},
\end{multline}
where $e^I_\mathrm{SNM}=e^I_\mathrm{B}(Y_p=1/2,T=0)$ and
$e^I_\mathrm{PNM}=e^I_\mathrm{B}(Y_p=0,T=0)$ are the baryon interacting energies
of the symmetric (SNM) and pure neutron matter (PNM), respectively, at
zero-temperature. 

Following~\cite{Burgio+Schulze.2010}, we assume that finite-temperature effects
do not modify the functional dependency of $f^I_\mathrm{B}$ on the proton fraction, i.e.
\begin{align}
f^I_\mathrm{B}(Y_p,T,n_\mathrm{B})={}&4Y_p(1-Y_p)f^I_\mathrm{SNM}(T,n_\mathrm{B})\notag\\
{}&+ (1-2Y_p)^2f^I_\mathrm{PNM}(T,n_\mathrm{B})\,, \label{eq:ffit} \end{align}
where $f^I_\mathrm{SNM}$ and $f^I_\mathrm{PNM}$ are the baryon interacting free
energies per baryon for symmetric and pure neutron matter. We verify the accuracy of this
assumption {\it a posteriori}: for \emph{given values} of temperature and baryon density, the difference between
the interacting baryon free energy and the quadratic fit 
in $Y_p$ is $\lesssim \unit[0.02]{MeV}$ for the GM3 EoS and $\lesssim \unit[0.05]{MeV}$ for the CBF-EI EoS,
to be compared with an interacting baryon free energy on the order of $\sim \unit[10]{MeV}$.

\begin{figure}
\includegraphics[width=\columnwidth]{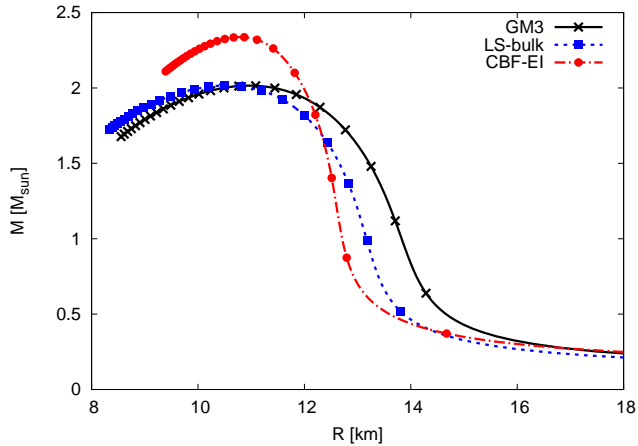}
\caption{$T=0$ mass-radius diagrams for the three EoSs considered
in this work.}
\label{fig:MR}
\end{figure}

We now discuss the dependency of the interacting part of the baryon free energy on the temperature and on the baryon
number density, i.e. the fitting formulae of symmetric and pure neutron matter, $f^I_\mathrm{SNM}(T,n_\mathrm{B})$ and
$f^I_\mathrm{PNM}(T,n_\mathrm{B})$, appearing in Eq.~\eqref{eq:ffit}. In the literature, there is no generally accepted
fitting formula for these
functions~\cite{Lattimer+Swesty.1991,Balberg+Gal.1997,Burgio+Schulze.2010,Ducoin+2011,Alam+2016}. In order to perform
our evolutionary numerical simulations, we need a fitting formula which is accurate in a wide density range, extending
from $n_\mathrm{B}\lesssim0.5$ fm$^{-3}$ to $n_\mathrm{B}\gtrsim0.001$ fm$^{-3}$ relevant for the core and the crust of
the star, respectively. Therefore, we can not use the fitting formula of~\cite{Burgio+Schulze.2010}, which is only
accurate for large densities. Moreover, we need a free energy with continuous second-order derivatives. Finally, the
following constraints have to be fulfilled: (i) $s\rightarrow0$ as $T\rightarrow0$; (ii) in the low density limit the
EoS must tend to that of a free gas, i.e., $f_\mathrm{B}^I\rightarrow0$, $s_\mathrm{B}^I\rightarrow 0$, and
$P_\mathrm{B}^I\rightarrow0$, as $n_\mathrm{B} \rightarrow 0$. Under these conditions, in the range
of temperatures and densities
considered (see Appendix~\ref{app:fit}), we find that a good trade-off
between number of parameters and precision of the fit is given by the following polynomial fitting
formula: \begin{multline}
\label{eq:fSNM_PNM}
f^I_{j}(n_\mathrm{B},T) = a_{1,j}n_\mathrm{B} + a_{2,j}n_\mathrm{B}^2 +
a_{3,j}n_\mathrm{B}^3+a_{4,j}n_\mathrm{B}^4 +\\
n_\mathrm{B}T^2(a_{5,j}+a_{6,j}T+a_{7,j}n_\mathrm{B} + a_{8,j}n_\mathrm{B}T),
\end{multline}
where $j=\{\mathrm{SNM};\mathrm{PNM}\}$.  We have performed the
fit~\eqref{eq:ffit}, \eqref{eq:fSNM_PNM} for the EoSs GM3 and CBF-EI. The
details of the fitting procedure, and the values of the coefficients $a_{n,i}$,
for these two EoSs are given in Appendix~\ref{app:fit}. For the LS-bulk EoS we
have used the analytical expression given
in~\cite{Lattimer+Swesty.1991},
\begin{equation} \label{eq:fLS}
f^I_\mathrm{B}=[a+4bY_p(1-Y_p)]n_\mathrm{B} + cn_\mathrm{B}^\delta -
Y_p\Delta_m\,, \end{equation} with \begin{align}
\delta={}&1.260,\notag\\
a={}&\unit[-711.0]{MeV\,fm^3},\notag\\
b={}&\unit[-107.1]{MeV\,fm^3},\notag\\
c={}&\unit[934.6]{MeV\,fm^{3\delta}},\notag\\
\Delta_m ={}& \unit[0]{MeV}\,.
\end{align}
This choice of parameters corresponds to a binding energy
$\mathrm{BE}=-\unit[16]{MeV}$, a saturation density
$n_s=\unit[0.155]{fm^{-3}}$, an incompressibility at saturation
$K_s=\unit[220]{MeV}$, a symmetry energy parameter at saturation
$S_v=\unit[29.3]{MeV}$, and a vanishing neutron-proton mass difference
$\Delta_m$.  As a comparison, the GM3 EoS has $n_s=\unit[0.153]{fm^{-3}}$,
$\mathrm{BE}=\unit[-16.3]{MeV}$, $K_s=\unit[240]{MeV}$, $S_v=\unit[32.5]{MeV}$,
and $\Delta_m=0$~\cite{Glendenning+Moszkowski.1991}; the CBF-EI EoS has
$n_s=\unit[0.16]{fm^{-3}}$, $\mathrm{BE}\simeq \unit[-11]{MeV}$,
$K_s=\unit[180]{MeV}$, $S_v=\unit[30]{MeV}$, and a vanishing \emph{bare}
neutron-proton mass difference (conversely, the proton and neutron effective
masses are different and change with density, temperature and composition).
In Fig.~\ref{fig:MR} we show the mass-radius diagram for cold
neutron stars for the three EoSs. To generate them, we have computed the zero-temperature EoSs at beta equilibrium,
considering both muons and electrons. The CBF-EI EoS has been linearly
extrapolated in the logarithms of $P$, $n_\mathrm{B}$, and $\epsilon$ for
densities higher than $n_\mathrm{B}=\unit[0.48]{fm^{-3}}$, enforcing causality
($c_s\le1$). This is necessary to describe the central region of stars with a
gravitational mass $M\gtrsim 1.64 M_\odot$, corresponding to a baryonic mass
$M_b\gtrsim 1.84 M_\odot$. The maximum mass for GM3 and LS-bulk is
$M_\mathrm{max}\simeq 2.02 M_\odot$, while for CBF-EI we get
$M_\mathrm{max}\simeq2.34 M_\odot$.

%

\subsection{Numerical implementation of the complete EoS} \label{ssec:eos_num}
%

In Sec.~\ref{ssec:fit} we discussed the interacting part of the baryon EoS (composed of
protons $p$ and neutrons $n$). In addition, for the baryon kinetic part and
for electrons and positrons we have
adopted the EoS of free fermions given
in~\cite{Eggleton+Faulkner+Flannery.1973,Johns+Ellis+Lattimer.1996}, and 
for the three neutrino families the EoS of free massless fermions given
in~\cite{Lattimer+Swesty.1991} [Eqns.~(C.1) and (C.3)]. The thermodynamical
quantities of the $i$-th lepton are given in terms of the temperature and of the
corresponding chemical potential $\mu_i$.

During the PNS evolution other particles are expected to appear, like hyperons,
muons, and tauons.
Since we are mostly interested in comparing how
 mean-field and  many-body EoSs affect the PNS evolution, we have focused on nucleons
(many-body EoSs have been developed mainly for nucleons).
Moreover, we do not include muons or tauons (as done also
in~\cite{Burrows+Lattimer.1986, Pons+.1999}), since a consistent treatment of
these particles would considerably increase the complexity of the transport
scheme.

The PNS structure and the transport equations (Sec.~\ref{ssec:equations}) suggest
to use as independent variables the pressure $P$, the entropy per baryon $s$,
and the electron lepton fraction $Y_L\equiv Y_e+Y_\nu$, where $Y_e\equiv
Y_{e^-}-Y_{e^+}$ and $Y_{\nu}=Y_{\nu_e}-Y_{\bar\nu_e}$. To determine the
different thermodynamical quantities of the complete EoS in terms of these
variables, we use a
Newton-Raphson cycle, in which we exploit the fitting formula discussed in
Sec.~\ref{ssec:fit} for the baryonic interacting quantities, along with the
leptonic EoS mentioned above, and we assume charge neutrality $Y_e\equiv
Y_{e^-}-Y_{e^+}=Y_p$, beta equilibrium
\begin{equation}
\label{eq:beta1}
\mu_{\nu_e}=\mu_p-\mu_n+\mu_{e^-}\,,
\end{equation}
and the requirement that muon and tau neutrinos are not trapped:
\begin{align}
\label{eq:no-trapped}
\mu_{\nu_\mu}={}&\mu_{\nu_\tau}=0,\\
\mu_{\bar \nu_{\{e,\mu,\tau\}}}={}&-\mu_{\nu_{\{e,\mu,\tau\}}}\,.
\end{align}

It is easy to obtain the GM3 quantities by directly solving the corresponding
mean-field equations.  For this reason, we have used GM3 as a benchmark for the
fitting procedure of the baryon free energy.  

\subsection{EoSs comparison}
\label{ssec:comparison}

\begin{figure*}
\includegraphics[width=\textwidth]{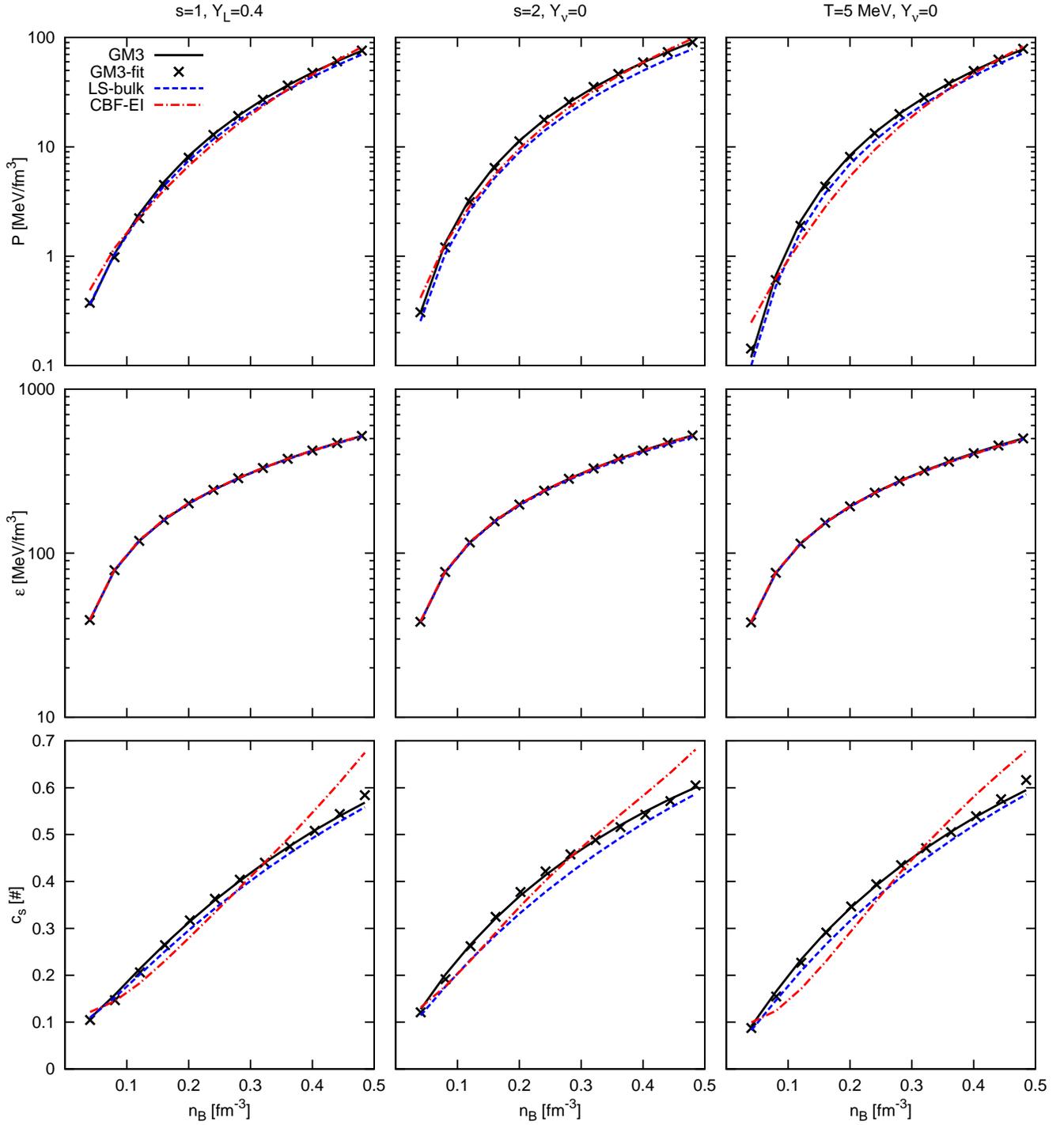}
\caption{The pressure (upper panels), the energy density (middle
panels), and  the speed of sound (bottom panels) are plotted
versus the baryon number density for the EoSs
considered in this paper, and for different values of selected parameters
(cases (i)-(iii) described in
Sec.~\ref{ssec:comparison}). The black solid line refers to the GM3 EoS
determined by solving numerically the mean-field equations, 
the black crosses to
the GM3 EoS determined through the fit and the procedure described in
Sec.~\ref{ssec:eos_num}, the blue dashed line to the LS-bulk EoS, and the red
dot-dashed line to the CBF-EI EoS.}
\label{fig:thermo_n}
\end{figure*}
\begin{figure*}
\includegraphics[width=\textwidth]{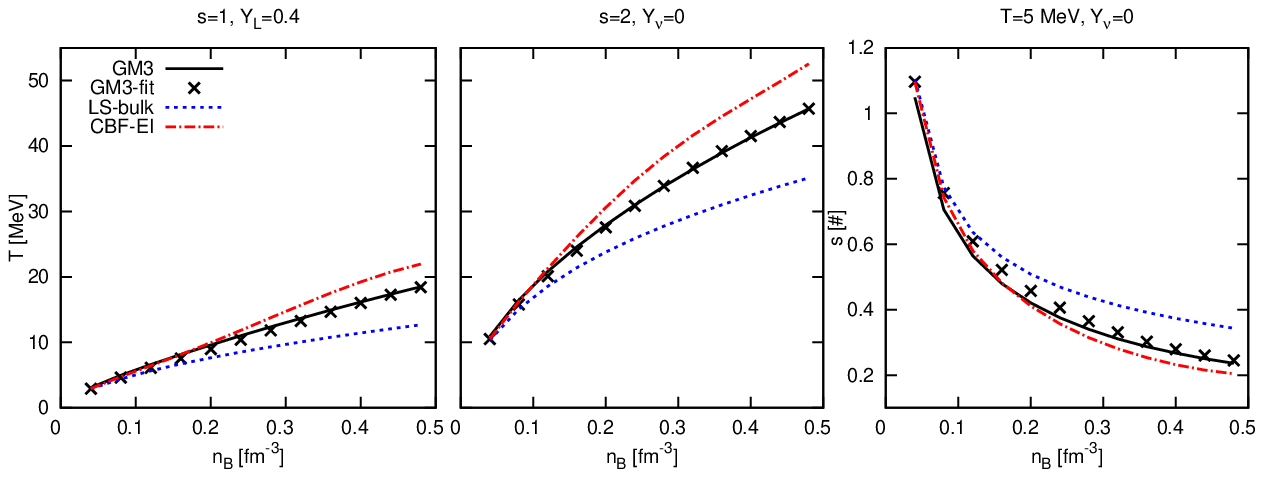}
\caption{Comparison among the three EoSs considered in this paper in the three cases described in
  Sec.~\ref{ssec:comparison}.
  In the left and central plots we show the temperature and in the right plot we show the entropy per baryon.
  Colors and line-styles are as in Fig.~\ref{fig:thermo_n}.
  }
\label{fig:fig2}
\end{figure*}

In this Subsection we compare the features of the three EoSs, and the 
accuracy of our fit for the baryon free energy,
by considering three cases: (i) $Y_L=0.4$ and $s=1$,
(ii)
$Y_\nu\equiv Y_{\nu_e}=0$ and $s=2$ (corresponding to the
end of the deleptonization phase), and (iii) $Y_\nu\equiv Y_{\nu_e}=0$ and
$T=\unit[5]{MeV}$ (which is the condition in most of the star at the end of our
simulations, i.e., toward the end of the cooling phase).

In Fig.~\ref{fig:thermo_n} we compare the behaviour of the EoS GM3
(continuous line), GM3-fit obtained using the fitting formula (crosses), LS-bulk (dashed line), and CBF-EI (dot-dashed line). We plot
the pressure, the energy density, and the sound speed, $c_s$, as functions of
the baryon number density, for the cases (i)-(iii) discussed above.
Fig.~\ref{fig:thermo_n} clearly shows that GM3-fit reproduces the behaviour of 
GM3 EoS.

As already noted by Pons {\it et al.}~\cite{Pons+.1999}, the pressure and the energy density in the three cases have a
similar dependence on the number density, since they mainly depend on the baryon interaction and degeneracy, rather than
on temperature. At the saturation density $n_s$ (whose exact value is slightly different for the three EoSs, but is in
the range $n_s=\unit[0.15\textrm{-}0.16]{fm^{-3}}$), the sound speed is slightly larger (lower) for the EoS with larger
(lower) incompressibility parameter $K_s$.  At high baryon density, the sound speed of the CBF-EI EoS is larger than
that of the LS-bulk and GM3 EoSs: this is due to a well-know problem of the many-body EoSs, which violate causality at
very high density. However, in the regime of interest for this paper, this unphysical behaviour can safely be neglected.

In Fig.~\ref{fig:fig2} we plot the temperature versus $n_B$ for $Y_L=0.4$ and
$s=1$, and $Y_\nu\equiv Y_{\nu_e}=0$ and $s=2$ (left and central panels) and the
entropy per baryon for $Y_\nu\equiv Y_{\nu_e}=0$ and $T=\unit[5]{MeV}$ (right
panel).  From the right panel we see that, at a fixed temperature, GM3 reaches a
given value of the entropy for a baryon density lower than that of LS-bulk and
higher than that of CBF-EI. This behaviour may be traced back to the fact that
particles in the GM3 EoS are less correlated than in the CBF-EI EoS, and more
correlated than in the LS-bulk EoS. Therefore, the CBF-EI describes a ``more
ordered'' nuclear matter than GM3 and the entropy is lower.
The left and central panels of  Fig.~\ref{fig:fig2}, where we plot the
temperature for fixed values of the entropy, show that CBF-EI is hotter than
GM3, which is hotter than LS-bulk.  As in Fig.~\ref{fig:thermo_n}, the GM3-fit
reproduces the behaviour of GM3 EoS.

\section{Neutrino diffusion coefficients}
%
\label{sec:diffusion}
\subsection{The equations}
The diffusion coefficients $D_2$, $D_3$, $D_4$ employed in the PNS evolution
(Sec.~\ref{sec:evolution}) are given in~\cite{Pons+.1999}:
\begin{align}
\label{eq:D2}
D_2={}&D_2^{\nu_e}+D_2^{\bar\nu_e},\\
\label{eq:D3}
D_3={}&D_3^{\nu_e}-D_3^{\bar\nu_e},\\
\label{eq:D4}
D_4={}&D_4^{\nu_e}+D_4^{\bar\nu_e}+4D_4^{\nu_\mu},\\
\label{eq:Dn}
D^{\nu_i}_n={}&\int_0^\infty\mathrm 
x^n\lambda_\mathrm{tot}^{\nu_i}(\omega)f^{\nu_i}(\omega)\big(1-f^{\nu_i}(\omega)\big)
dx,\\
\label{eq:lambdatot}
\lambda_\mathrm{tot}^{\nu_i}(\omega)={}& \left(\sum_{j\in\mathrm{reactions}} \frac{\sigma_j^{\nu_i}(\omega)}{V}\right)^{-1},
\end{align}
where $f^{\nu_i}(\omega)=[1+\exp((\omega-\mu_{\nu_i})/T)]^{-1}$ and $\lambda_\mathrm{tot}^{\nu_i}(\omega)$ are the
distribution function\footnote{Here we set $\hbar=1$ and assume that integrals are normalized
as in~\cite{Reddy+Prakash+Lattimer.1998, Pons+.1999}.} and
the total mean free path of a $\nu_i$ neutrino of energy $\omega$,
respectively, and $x=\omega/T$. The $\nu_i$ neutrino cross
section of the $j$-th reaction is denoted with $\sigma_j^{\nu_i}$. All 
quantities depend upon the temperature and the particle chemical potentials,
which are determined by the underlying EoS.

To determine the $\sigma_j^{\nu_i}$ we adopt the mean-field approach
of~\cite{Reddy+Prakash+Lattimer.1998} [Eq.~(82)] that accounts for in-medium effects,
including the scattering of
all neutrino types on electrons, protons, and neutrons, and the absorption of
electron neutrinos and electron anti-neutrinos on neutrons and protons,
respectively, with the corresponding inverse processes, i.e.
\begin{align}
\nu_i + n \rightleftharpoons{}& \nu_i + n,\\
\nu_i + p \rightleftharpoons{}& \nu_i + p,\\
\nu_i + e^- \rightleftharpoons{}& \nu_i + e^-,\\
\nu_e + n \rightleftharpoons{}& e^- + p,\\
\bar \nu_e + p \rightleftharpoons{}& e^+ + n.
\end{align}
Furthermore, we assume that the cross-sections of all non-electronic neutrinos
coincide with that of muon neutrinos.
We do not include nucleon-nucleon Bremsstrahlung~\cite{Fischer.2016}.

\subsection{Effective masses and single-particle potentials}
%
\label{ssec:mstar}
Identifying single-particle properties in interacting
systems involves non trivial conceptual difficulties. However, due to translation
invariance, in infinite matter single-particle states are labeled by the
momentum ${\bf k}$, and the corresponding spectrum can be unambiguously
identified.  Within  the non-relativistic many body theory, the spectrum of an
interacting particle can be expressed as 
\begin{equation} \mathcal E(k) = m +
\frac{k^2}{2m} + U (k)\, , \end{equation} where $k=|{\bf k}|$ and $U(k)$ is the
momentum-dependent single-particle potential.

A widely used parametrization of $\mathcal E(k)$ is given in terms of 
momentum-independent 
effective mass $m^\ast$ and  single-particle potential $U$ 
\begin{equation}
\label{eq:effSpectrum}
\mathcal E(k) \simeq  m+ \frac {k^2} {2m^\ast} + U
\end{equation}

Since the baryonic contributions to the mean free paths and diffusion coefficients are 
mostly given by particles whose energies are close to the particle chemical potential, it is convenient 
to determine $m^\ast$ and $U$ from the behaviour of the spectrum near the Fermi momentum
\begin{align}
\label{eq:mastNR}
\frac 1{m^\ast_i}={}&\frac1{k_F}\frac{\partial \mathcal
E_i}{\partial k}(k_F),\\
\label{eq:UNR}
U_i={}&\mathcal E_i(k_F) -\frac{k_F^2}{2m^\ast_i} - m,
\end{align}
where $i=(\{p;n\},Y_p,T,n_b)$. 

Within the CBF-EI approach $\mathcal E(k)$ has been obtained at the Hartree-Fock level
using the same effective potential employed for the calculation of the EoS. Therefore, the
effective masses and the single-particle potentials are \emph{consistent} with the EoS.

Eq. (\ref{eq:effSpectrum}) can be easily generalized to the relativistic case 
\begin{equation}
\label{eq:Estar}
\mathcal E(k) = \sqrt{k^2+m^\ast{}^2} + U^\ast\, ,
\end{equation}
where we have introduced $U^\ast=U-m^\ast+m$.  To treat the neutrino transport
for the CBF-EI EoS consistently with that of the GM3 and LS-bulk EoSs, 
we compute the neutrino diffusion coefficients using Eq.~\eqref{eq:Estar} 
and the effective masses and single-particle potentials given in 
Eqs.~\eqref{eq:mastNR} and \eqref{eq:UNR}. 

We have verified that this approach reproduces, for the CBF-EI EoS, the correct
baryon densities within $\sim10\%$ at saturation density\footnote{The other baryon EoS quantities cannot be recovered from the
baryonic spectrum as one should account also for the meson contributions. This
is true also for the GM3 EoS, for which the description in term of effective spectrum is
exact.}. This is important, because the neutrino mean
free path [Eq.~\eqref{eq:lambdatot}] is an ``intensive'' quantity, and it
depends on the baryon distribution functions.  A discrepancy on the baryon
densities $n_\mathrm{B}$ and the proton fraction $Y_p$ would yield diffusion
coefficients computed at wrong values of $n_\mathrm{B}$ and $Y_p$.

This applies also to the LS-bulk EoS, for which we have assumed that the baryon
effective masses are equal to the neutron bare mass. To satisfy the
aforementioned constraint (that the effective spectrum description yields the
correct baryon density and proton fraction), we use a non
vanishing single-particle potential given by
\begin{equation}
U_i^\ast=\mu_i^I=\mu_i-\mu_i^K,
\end{equation}
where $i=\{p;n\}$, $\mu$ is the chemical potential, and $\mu^I$ and $\mu^K$ are
the interacting and free part of the particle chemical potential, respectively.

\subsection{Numerical implementation}
%
\label{ssec:diff_num}

The neutrino diffusion coefficients are evaluated in the PNS evolution code by
linear interpolation of a three-dimensional table, evenly spaced in $Y_\nu$ (the
neutrino number fraction), $T$, and $n_\mathrm{B}$. The table has been produced
consistently with the underlying EoS, in the following way.  We have first
solved the EoS using the method described in Sec.~\ref{ssec:eos_num}, obtaining
the proton fraction $Y_p$ as function of $Y_\nu$, $T$, and
$n_\mathrm{B}$. The proton and neutron chemical potentials, effective masses, and the single-particle
potentials for the GM3-fit and CBF-EI EoSs have
been obtained by linear interpolation of a table evenly spaced in $Y_p$, $T$,
and $n_\mathrm{B}$. From these quantities, we determine the neutrino
cross-sections, and finally the neutrino diffusion coefficients
[Eqs.~\eqref{eq:D2}-\eqref{eq:Dn}].

\subsection{EoS comparison}
\label{ssec:diff_comparison}

\begin{figure*}
\includegraphics[width=\textwidth]{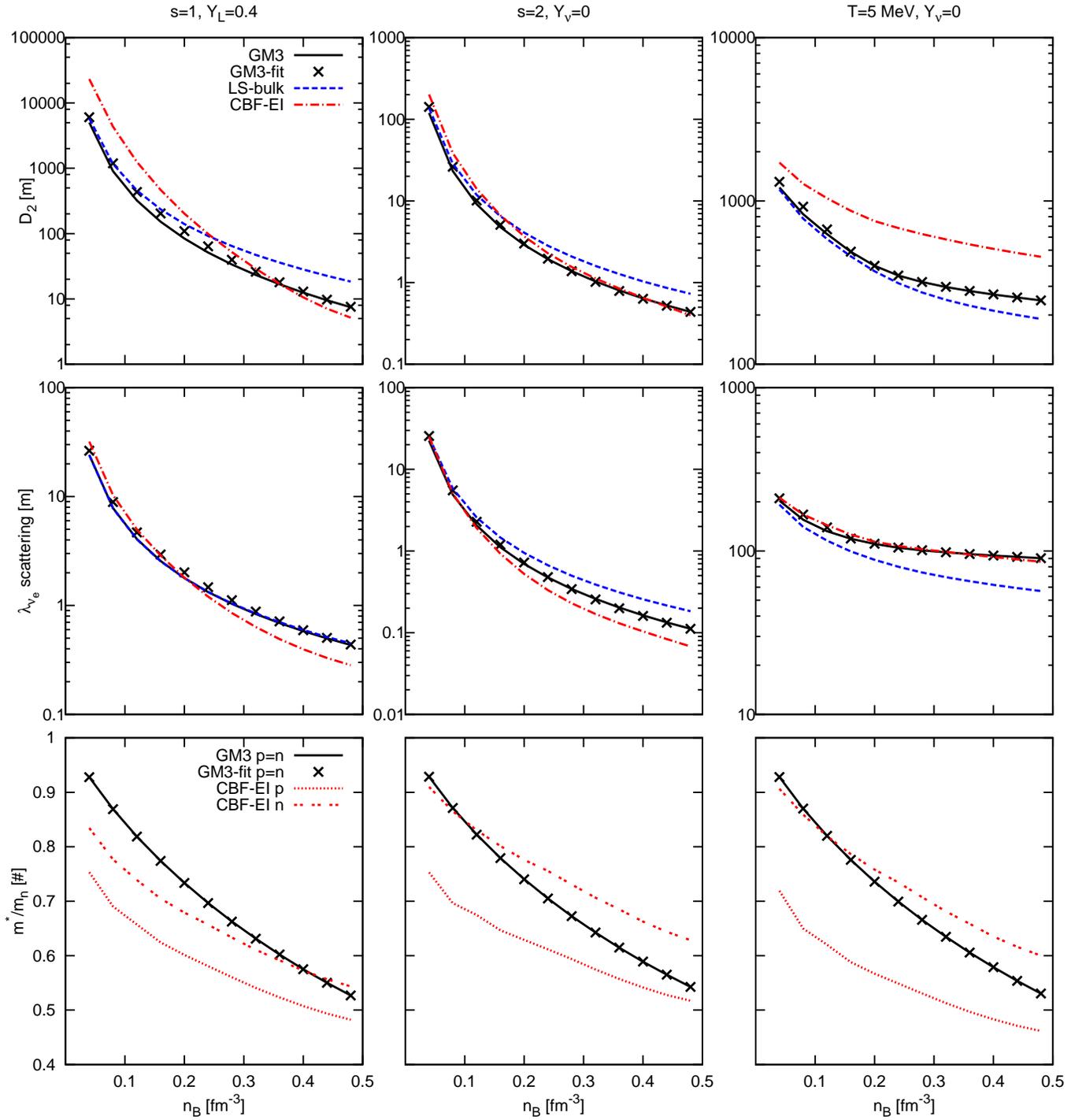}
\caption{Comparison between the diffusion coefficient $D_2$ (upper panel), the
electron neutrino scattering mean free paths [middle panel,
the neutrino incoming energy is $E_{\nu_e}=\max(\mu_{\nu_e},\pi T)$], and 
the baryon effective masses, $m^\ast/m_n$ (lower panel) for the three EoSs 
considered in this paper in the three cases described in
Sec.~\ref{ssec:comparison}. 
For the GM3 EoS, the effective masses of proton and neutron are
identical~\cite{Glendenning.1985,Glendenning+Moszkowski.1991}.  We do not show the LS-bulk EoS effective masses, since
we have set them equal to the bare ones, $m^\ast_{\{p,n\}}/m_n=1$.
Colors and line-styles are as in
Fig.~\ref{fig:thermo_n},
apart for the line-styles in the lower panel, where the CBF-EI proton (neutron) effective masses are dotted (double-dashed).}
\label{fig:diff}
\end{figure*}

In Fig~\ref{fig:diff} we plot the neutrino diffusion coefficient $D_2$, the
electron neutrino scattering mean free path, and the baryon effective masses in the three cases
described in Sec.~\ref{ssec:comparison}.  The incident neutrino energy which we
have used to compute the neutrino mean free path is
$E_{\nu_e}=\max(\mu_{\nu_e},\pi T)$. To understand the role of the interactions and
of finite temperature in the neutrino diffusion, we consider their effects on the
baryon distribution function.  To fix ideas, let us  consider the 
distribution function of a non-relativistic fermion gas,
\begin{equation}
\label{eq:fake_distr_func} f(k)=\frac1{h^3}\left(1+\mathrm e^{\frac
{k^2}{2m^\ast T} - \frac{\mu-U-m}{T}}\right)^{-1},
\end{equation}
If one decreases the temperature $T$ or effective mass $m^\ast$,
$f$ approaches a Heaviside function, whereas increasing $T$ or $m^\ast$
it becomes smoother. Because of the Pauli principle, at lower temperatures $T$ and effective masses
$m^\ast$ lower energy neutrinos can interact only with particles
near the Fermi sphere, and therefore the mean free paths and diffusion
coefficients increase. Conversely, a greater temperature and effective mass
imply that
the mean free paths and the diffusion coefficients are smaller.  The scattering
mean free paths  reflect the temperature dependence of the three EoSs: when the matter is
hotter, the scattering is more effective (cf.~lower plots of
Fig.~\ref{fig:fig2}). At equal temperature, the interaction is more
effective when the effective mass is greater.
The behavior of the diffusion coefficient $D_2$ results from a complex interplay between
scattering and absorption, for which the effective masses and single particle
potentials play an important role.
The comparison between the diffusion coefficient $D_2$ for the three EoSs
suggests that towards the end of the cooling phase (in which the thermodynamical conditions are roughly similar to those in
the right plots of Fig.~\ref{fig:diff}), the CBF-EI star evolves faster than the other EoSs.

As in Figs.~\ref{fig:thermo_n} and \ref{fig:fig2}, GM3-fit (for which the baryon spectra effective parameters are
determined by table interpolation, Sec.~\ref{ssec:diff_num}) reproduces the results
of the GM3 EoS.

%

%
\section{PNS evolution}
\label{sec:evolution}
%

\subsection{The equations} \label{ssec:equations}
%
We developed a numerical code to model the PNS evolution. Our code is similar
to that of Pons {\it et al.}~\cite{Pons+.1999}: it is energy averaged (the neutrino distribution
function has been assumed Fermi-Dirac and in thermal equilibrium with matter),
general relativistic (we include GR consistently both in the stellar structure
and in the neutrino transport), spherically symmetric (the stellar structure is
determined by integrating the TOV equations), and flux limited (we use the
diffusion approximation and apply a flux limiter to preserve causality in the
optically thin regions near the border). Since we want to focus
on how the EoS affects the evolution and the gravitational wave emission, we do
not include convection in our simulations (see e.g.{}\cite{Roberts+2012} for
a PNS simulation including convection with the mixing length theory) nor
accretion~\cite{Burrows.1988, Takiwaki+Kotake+Suwa.2014,
Melson+Janka+Marek.2015, Muller.2015}, that are both present in
this phase.
The spacetime metric is
\begin{equation}
\label{eq:metric}
\mathrm ds^2=-\mathrm e^{2\phi}\mathrm dt^2 + \mathrm e^{2\lambda} \mathrm dr^2
+ r^2 \mathrm d\Omega,
\end{equation}
where $\phi$ and $\lambda$ are metric functions that depend on the radius $r$,
$t$ is the time for an observer at infinity, and $\mathrm d\Omega$ is the
 element of solid angle.

The stellar structure, at each timestep, is given by the TOV equations,
\begin{align}
\label{eq:TOV1}
\frac{\mathrm d r}{\mathrm d a} ={}&\frac{1}{4\pi r^2n_\mathrm{B}\mathrm e^\lambda},\\
\label{eq:TOV2}
\frac{\mathrm d m}{\mathrm d a} ={}& \frac {\epsilon}{n_\mathrm{B} \mathrm e^\lambda},\\
\label{eq:TOV3}
\frac{\mathrm d \phi}{\mathrm d a}={}& \frac{\mathrm e^\lambda}{4\pi r^4 n_\mathrm{B}}(m + 4\pi r^3 P),\\
\label{eq:TOV4}
\frac{\mathrm d P}{\mathrm d a}={}&- (\epsilon +P)\frac{\mathrm e^\lambda}{4\pi r^4n_\mathrm{B}}(m+4\pi r^3P),
\end{align}
where $r$ is the radius, $m$ is the gravitational mass at radius $r$, $a$ is the enclosed baryon number at
radius $r$, $\epsilon$ is the total energy density (matter \emph{plus} neutrino energy density),
and the metric function $\lambda$ is given by
\begin{equation}
\label{eq:lambda}
\mathrm e^{-\lambda}=\sqrt{1-\frac{2m}{r}}.
\end{equation}

The neutrino diffusion equations are~\cite{Lindquist.1966,Pons+.1999}
\begin{align}
\label{eq:Fnu}
F_\nu=-\frac{\mathrm e^{-\lambda}\mathrm e^{-\phi}T^2}{6\pi\hbar^3}
\left(D_3\frac{\partial(T\mathrm e^\phi)}{\partial r}+
(T\mathrm e^\phi)D_2\frac{\partial \eta}{\partial r}\right),&\\
\label{eq:Hnu}
H_\nu=-\frac{\mathrm e^{-\lambda}\mathrm e^{-\phi}T^3}{6\pi\hbar^3}
\left(D_4\frac{\partial(T\mathrm e^\phi)}{\partial r}+
(T\mathrm e^\phi)D_3\frac{\partial \eta}{\partial r}\right),&\\
\label{eq:dYldt}
\frac{\partial Y_L}{\partial t}+\frac{\partial (\mathrm e^\phi4\pi r^2F_\nu)}{\partial a}=0,&\\
\label{eq:dsdt}
T\frac{\partial s}{\partial t}+\mu_{\nu_e}\frac{\partial Y_{L}}{\partial t}+\mathrm e^{-\phi}
\frac{(e^{2\phi}4\pi r^2H_\nu)}{\partial a}=0,&
\end{align}
where $F_\nu$ and $H_\nu$ are the neutrino number and energy\footnote{Eq.~\eqref{eq:dsdt}
is derived from the sum of the transport equations for the neutrino and matter energy,
\begin{align}
\frac{\partial e_\nu}{\partial t}-\frac{P_\nu}{n_\mathrm{B}}
\frac{\partial n_\mathrm{B}}{\partial t}+\mathrm e^{-\phi}
\frac{(e^{2\phi}4\pi r^2H_\nu)}{\partial a}={}&+\mathrm e^{\phi}\frac{S_E}{n_\mathrm{B}},\\
\frac{\partial e_{matter}}{\partial t}-\frac{P_{matter}}{n_\mathrm{B}}
\frac{\partial n_\mathrm{B}}{\partial t}={}&-\mathrm e^{\phi}\frac{S_E}{n_\mathrm{B}},
\end{align}
where $S_E$ is the energy and momentum integrated source term for the energy~\cite{Pons+.1999}.} fluxes, respectively,
$\eta=\mu_{\nu_e}/T$ is the electron neutrino degeneracy, and the diffusion coefficients
$D_2$, $D_3$, and $D_4$ are given by Eqs.~\eqref{eq:D2}, \eqref{eq:D3}, and
\eqref{eq:D4}. $Y_L\equiv Y_e+Y_\nu=Y_{e^-}+Y_{\nu_e}-Y_{e^+}-Y_{\bar\nu_e}$ is the
\emph{total} electron lepton fraction.

Previous PNS studies have found that the beta equilibrium does occur almost everywhere in the star during the
evolution~\cite{Burrows+Lattimer.1986,Pons+.1999}.  Therefore, to additionally simplify the equations, we enforce beta
equilibrium (Eq.~\eqref{eq:beta1}, as in~\cite{Keil+Janka.1995}).  We have checked \emph{a posteriori} that beta
equilibrium is respected almost everywhere in the star during the evolution, apart for a thin region near the stellar
surface at early times (see Appendix~\ref{ssec:beta-equilibrium}).

\subsection{Numerical implementation}
\label{ssec:evolution_implementation}

In a PNS in beta equilibrium, all thermodynamical quantities can be uniquely
determined in terms of three independent variables. A natural choice, looking
at the evolution and structure equations, is to use as independent variables
the pressure $P$, the entropy per baryon $s$, and the lepton fraction $Y_L$
(see Sec.~\ref{sec:EOS}).

We started the simulation assuming entropy and lepton fraction initial profiles
similar to those of~\citet{Pons+.1999} (see Fig.~\ref{fig:profiles}), that is, the profiles obtained
in~\citet{Wilson+Mayle.1989} at the end of their core-collapse simulation
(200~ms after core bounce). The entropy and lepton fraction
content of the PNS depend on the stellar mass. To qualitatively reproduce this
behaviour, we have rescaled the entropy and lepton fraction profiles with the
stellar baryon mass $M_\mathrm{B}$,
\begin{align}
s(a,t=\unit[200]{ms})={}&\frac{M_\mathrm{B}}{M'_\mathrm{B}}s'\!\left(a',t=\unit[200]{ms}\right),\\
Y_L(a,t=\unit[200]{ms})={}&\frac{M_\mathrm{B}}{M'_\mathrm{B}}Y_L'\!\left(a',t=\unit[200]{ms}\right),\\
a={}&\frac{M_\mathrm{B}}{M'_\mathrm{B}}a',
\end{align}
where the prime refers to the reference profiles of~\citet{Wilson+Mayle.1989}.
Using these initial entropy and lepton fraction profiles at 200~ms, we have first
determined the initial structure of the star solving the TOV Eqs.~\eqref{eq:TOV1}--\eqref{eq:TOV4} by numerical
relaxation (\cite{Press+book}, Sec.~17.3).  We have then evolved the star solving separately the structure and diffusion
equations in a series of iterative predictor-corrector steps, as in~\cite{Pons+.1999}.  To prevent superluminar fluxes,
the neutrino number and energy fluxes [Eqs.~\eqref{eq:Fnu} and \eqref{eq:Hnu}] have been numerically limited using the
flux limiter of Levermore and Pomraning~\cite{Levermore+Pomraning1981}, which is relevant
near the stellar surface, where the matter is optically thin to neutrinos and the diffusion approximation breaks down.

We discuss the numerical convergence of our code in
Appendix~\ref{ssec:conservation}. More details on the code are reported
in~\cite{Camelio:thesis}.

\subsection{Results}
\label{ssec:evolution}

We now discuss how the PNS evolution depends on the EoS adopted and the total
stellar baryon mass. In Fig.~\ref{fig:central_1.60} we show the evolution of
the central and maximum temperature, central entropy per baryon, central
neutrino and proton fraction, and central baryon density
for the three nucleonic EoSs considered in this paper and for the total baryon
mass $M_\mathrm{B}=\unit[1.60]{M_\odot}$.  It is apparent that the evolution
with the GM3-fit EoS reproduces that with the GM3 EoS. Therefore, in the rest
of this paper we do not distinguish between the GM3-fit and GM3 EoS.  In
Tab.~\ref{tab:central} we summarize the timescales of the evolutionary phases
and the maximum central temperature for the three EoSs and for three stellar
baryon masses $M_\mathrm{B}=\unit[(1.25,1.40,1.60)]{M_\odot}$.
In Fig.~\ref{fig:L_M_R} we plot the time dependency of the total neutrino
luminosity, gravitational mass, and stellar radius of a PNS evolved using the
three EoSs discussed in this paper and with the stellar baryon masses
$M_\mathrm{B}=\unit[(1.25,1.40,1.60)]{M_\odot}$.

\begin{figure*}
\includegraphics[width=\textwidth]{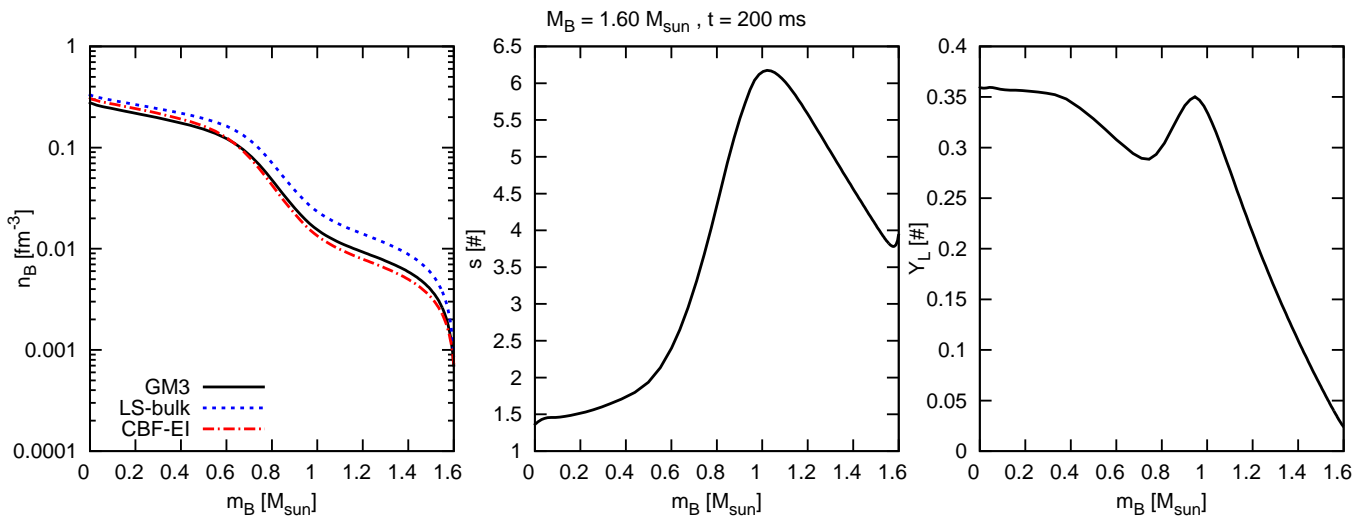}
\caption{The initial profiles (at $t=\unit[200]{ms}$) of the baryon density (left),
entropy per baryon (center), and lepton fraction (right) are plotted versus the
enclosed baryonic mass for a
$M_\mathrm{B}=\unit[1.60]{M_\odot}$ star.
The \emph{initial} entropy per baryon and lepton fraction profiles are the same
for the three EoSs adopted, whereas the baryon density depends on the EoS
(colors and line-styles are as in Fig.~\ref{fig:thermo_n}).}
\label{fig:profiles}
\end{figure*}

\begin{figure*}
\includegraphics[width=\textwidth]{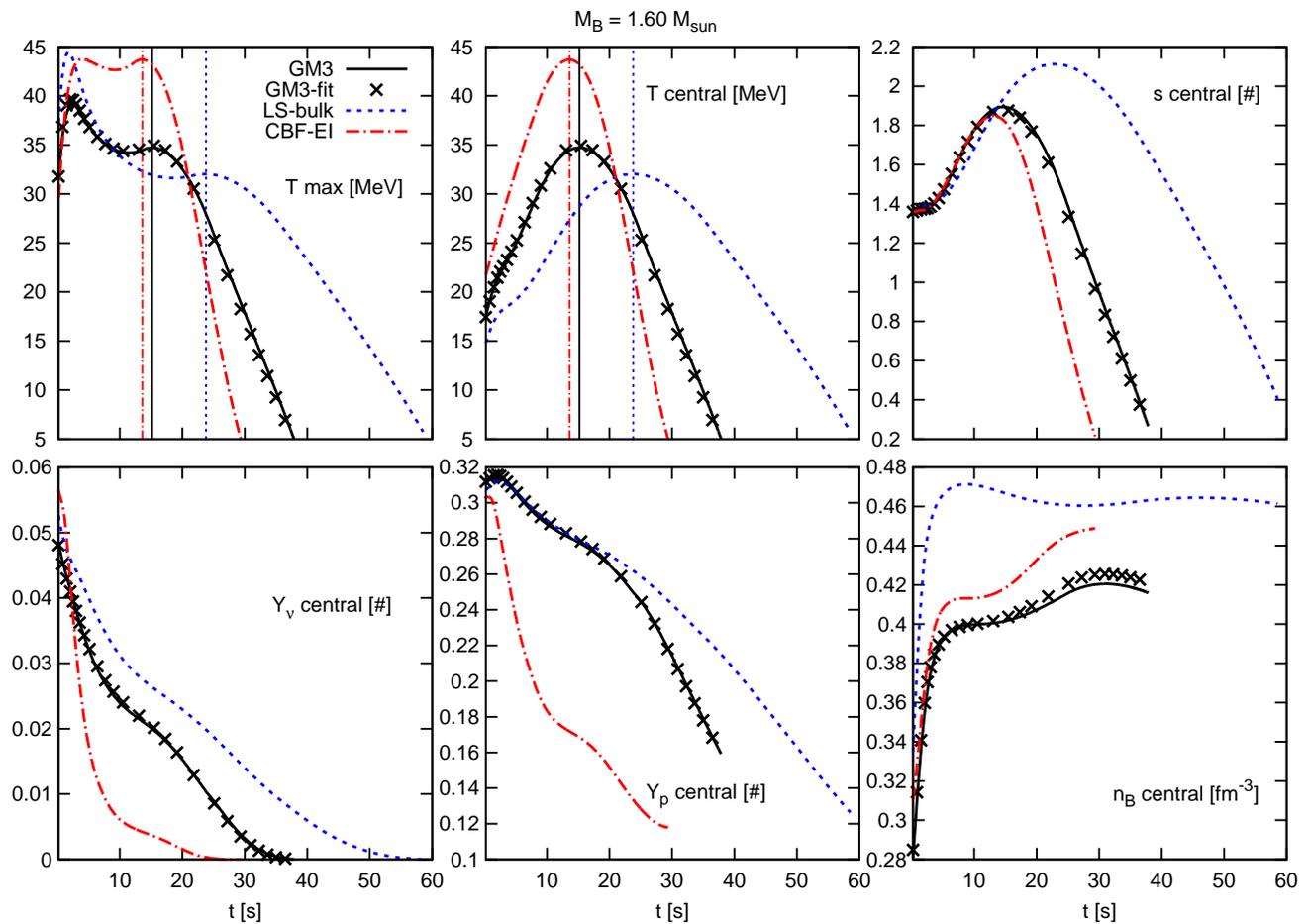}
\caption{Time dependence of the maximum and central temperature (left and
central upper plots), central entropy per baryon (right upper plot), neutrino
and proton fraction (left and central lower plots) and central baryon
density (right lower plot), for a star with total baryon mass
$M_\mathrm{B}=\unit[1.60]{M_\odot}$ evolved using the three EoSs. Colors and
line-styles are as in Fig.~\ref{fig:thermo_n}. The three vertical lines in the temperature plots
mark the end of the Joule-heating phase (see text).}
\label{fig:central_1.60}
\end{figure*}

\begin{figure*}
\includegraphics[width=\textwidth]{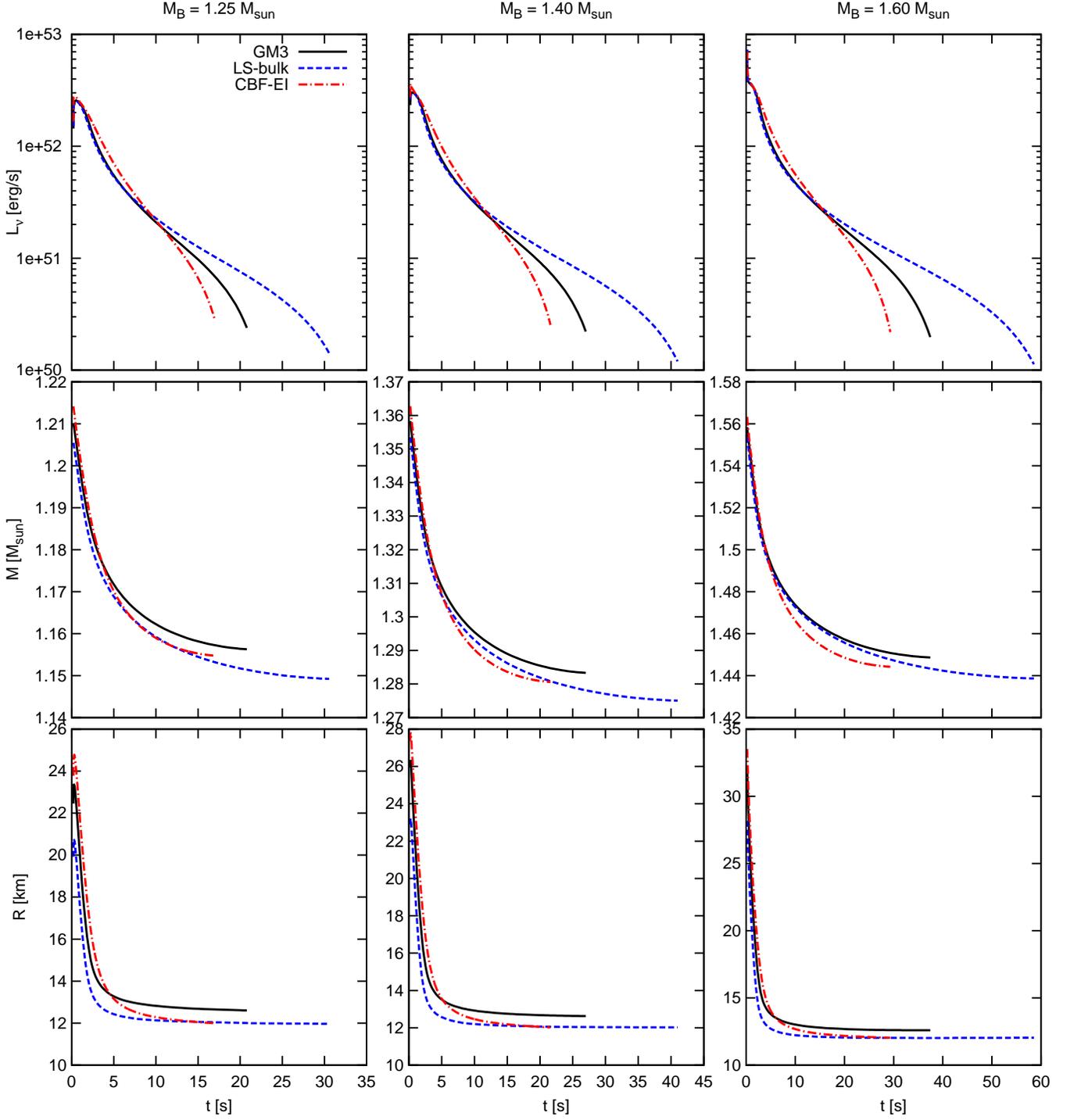}
\caption{Time dependence of the total neutrino luminosity (upper panels),
gravitational mass (middle panels), and stellar radius (lower panels) of a PNS
evolved with the three EoSs considered in this paper and the baryon stellar
masses $M_\mathrm{B}=\unit[(1.25,1.40,1.60)]{M_\odot}$.  The black solid lines
correspond to the GM3 EoS determined through the fit and the procedure
described in Sec.~\ref{ssec:eos_num}, the blue dashed lines to the LS-bulk EoS,
and the dot-dashed red lines to the CBF-EI EoS. The gravitational masses at
the end of the simulations are: for $M_\mathrm{B}=1.25M_\odot$,
$M_\mathrm{GM3}=1.1554 M_\odot$, $M_\mathrm{LS-bulk}=1.1492 M_\odot$,
$M_\mathrm{CBF-EI}=1.1548M_\odot$;
 for $M_\mathrm{B}=1.40M_\odot$,
$M_\mathrm{GM3}=1.2824 M_\odot$, $M_\mathrm{LS-bulk}=1.2750 M_\odot$,
$M_\mathrm{CBF-EI}=1.2806M_\odot$;
and  for $M_\mathrm{B}=1.60M_\odot$,
$M_\mathrm{GM3}=1.4478 M_\odot$, $M_\mathrm{LS-bulk}=1.4386 M_\odot$,
$M_\mathrm{CBF-EI}=1.4442M_\odot$.
}
\label{fig:L_M_R}
\end{figure*}

\begin{table}
\caption{Significant quantities describing the PNS evolution for the three EoSs
described in this paper and for three stellar baryon masses. The first column
contains the name of the EoS, the second column contains the stellar baryon
mass, the third and fourth columns contain the maximum central temperature and the
corresponding time (the latter approximately corresponds to the end of the
Joule-heating phase), respectively, the fifth column contains the time at which the central
neutrino fraction becomes equal to $Y_\nu=0.005$ (this is an indication on the
duration of the deleptonization phase), and the sixth column contains the time
at which our simulation ends (namely, when the central temperature becomes
equal to $T=\unit[5]{MeV}$). All simulations start at $t_\mathrm{start}=\unit[0.2]{s}$.}
\label{tab:central}
\centering
\begin{tabular}{cccccc}
EOS & $M_\mathrm{B}$ [$M_\odot$]& $T_\mathrm{max}$ [MeV] & $t_\mathrm{Jh}$ [s] & $t_\mathrm{del}$ [s] & $t_\mathrm{end}$ [s] \\
\hline
GM3     & 1.25 & 24.6 &  9.0 & 13.1 & 20.8 \\
GM3     & 1.40 & 28.7 & 11.2 & 18.6 & 27.1 \\
GM3     & 1.60 & 34.9 & 15.2 & 27.9 & 37.7 \\
LS-bulk & 1.25 & 23.6 & 13.5 & 17.5 & 30.6 \\
LS-bulk & 1.40 & 26.6 & 17.6 & 26.3 & 41.0 \\
LS-bulk & 1.60 & 32.1 & 23.8 & 41.4 & 59.2 \\
CBF-EI  & 1.25 & 32.3 & 7.31 & 3.46 & 17.0 \\
CBF-EI  & 1.40 & 37.0 & 9.55 & 5.65 & 21.6 \\
CBF-EI  & 1.60 & 43.7 & 13.6 & 11.7 & 29.4
\end{tabular}
\end{table}

The qualitative behaviour of the stellar evolution is the same for the three
EoSs and the three stellar masses, even though the timescales and the
thermodynamical profiles are quantitatively different
(Fig.~\ref{fig:central_1.60} and Tab.~\ref{tab:central}).  At the beginning of
the evolution, which is 200 ms from core bounce, the PNS has a (relatively) low
entropy core and a high entropy envelope (see
Fig.~\ref{fig:profiles}).  The neutrino chemical potential initially is very
high in the center of the star; the process of neutrino diffusion transfers this
degeneracy energy from neutrinos to the matter and this causes the heating of
the PNS core. Moreover, on timescales of about 10 s, the star contracts from
about $30$ km to its final radius of about $12$--$13$ km. The region which is
affected the most from this contraction is the envelope, whose temperature
significantly increases.  At the same time, the steep negative neutrino chemical
potential gradient in the envelope causes a deleptonization of the envelope.
The neutrinos leave the star, bringing with them energy. The joint effect of the
envelope heating caused by contraction and the cooling caused by neutrino
emission is apparent in the behaviour of the maximum stellar temperature: before
the central temperature, $T_c$, reaches its maximum, the maximum temperature
reached  in the interior of the star, $T_{max}$,
increases, reaches a maximum value, and then decreases
(Fig.~\ref{fig:central_1.60}).
The initial phase, during which the central temperature increases, lasts for
several seconds and has been referred to as \emph{Joule heating} phase in
previous works \citep{Burrows+Lattimer.1986, Keil+Janka.1995}. We may place the
end of this phase approximately at the time $T_c$ reaches its maximum
(vertical dotted lines in Fig.~\ref{fig:central_1.60}); 
at that time the central temperature is also the maximum
stellar temperature (see Fig.~\ref{fig:central_1.60}).

After the Joule
heating phase, there is a general cooling of the star as the deleptonization
proceeds.  In~\cite{Burrows+Lattimer.1986,Keil+Janka.1995} it was found that
the end of the Joule heating phase coincides with the end of deleptonization,
whereas in~\cite{Pons+.1999}, with the GM3 EoS and a more refined treatment
of neutrino opacities, it was found that the deleptonization is longer than the
Joule-heating phase.  We agree with this last result
 for the stars with the GM3 and LS-bulk EoSs, whereas in the case
of the CBF-EI EoS we find that most of neutrinos have been radiated away by the end
of the Joule-heating phase
(Fig.~\ref{fig:central_1.60} and Tab.~\ref{tab:central}).

Our results for the $M_\mathrm{B}=\unit[1.60]{M_\odot}$ PNS with the GM3 EoS are
in qualitative agreement with those of~\cite{Pons+.1999}. In particular, the
duration of the Joule-heating phase is in good agreement (cf.{}
Fig.~\ref{fig:central_1.60} of this paper with Fig.~17 in~\cite{Pons+.1999});
however we find lower stellar temperatures and a shorter cooling phase.

We think that the quantitative differences\footnote{The differences amount
in about $10\%$ in the value of the central temperature maximum
and of the deleptonization time, and in less than
$2\%$ for the time of the end of Joule-heating phase, compare Tab.~\ref{tab:central}
and Fig.~17 of \citealp{Pons+.1999}.} between
our results and those of~\cite{Pons+.1999} are due to differences in the initial
profiles and in the details of the treatment of the diffusion processes.

For each EoS, the evolutionary timescales are smaller for stars with smaller 
baryonic mass, see Tab.~\ref{tab:central}. This is due to the way we have rescaled
the initial entropy per baryon and lepton fraction profiles with  $M_B$,
but also to the fact that a lower stellar mass corresponds
to lower baryonic densities and then to longer neutrino mean free paths.
We also notice that a lower stellar mass corresponds to lower temperatures. This
again depends on the initial entropy profiles and on the different densities
present in the star, see Fig.~\ref{fig:fig2}: at a given entropy per baryon and
lepton (or neutrino) fraction, lower densities (i.e. lower masses) 
correspond to lower temperatures.
To simulate a fully consistent PNS evolution, one should use 
initial profiles generated by core-collapse simulations of  stars with the same
baryonic mass
(see~\cite{Pons+.1999} for a study on how the initial conditions affect the PNS
evolution).

Fig.~\ref{fig:fig2} shows that, at fixed entropy, CBF-EI EoS
is hotter than the GM3 EoS, which is hotter than the LS-bulk EoS (see
discussion in Sec.~\ref{ssec:comparison}).  Since, for a given stellar mass,
the initial entropy profiles are the same for the three EoSs, then the CBF-EI
star reaches temperatures higher than the GM3 stars, which in turn reaches
temperatures higher than the LS-bulk star, see Fig.~\ref{fig:central_1.60} and
Tab.~\ref{tab:central}.

The fact that the LS-bulk evolution is slower than the GM3 one, which in turn is
slower than that of the CBF-EI EoS, may well be explained by the fact that in
the many-body CBF-EI EoS nuclear correlations are stronger than in the
mean-field GM3 EoS, in which in turn are stronger than in the LS-bulk EoS (where
the baryon masses are equal to the bare ones).  A
smaller neutrino cross section is a consequence of a greater baryon correlation
(Sec.~\ref{ssec:diff_comparison}).  This effect is relevant even at the
mean-field level, where one adopts the description of the baryon spectra in term
of effective masses and single-particle potentials to obtain the diffusion
coefficients. For example, the fact that the proton effective mass is
significantly smaller than the neutron one in the CBF-EI framework is a
consequence of the tensor correlations which are stronger in the n-p channel
than in the n-n or p-p channels.

To check this interpretation, that is, that the different timescales are mainly
due to the details of the microphysics (i.e., the baryon spectra and hence the
neutrino mean free paths and diffusion coefficients), we have run a simulation
of a $M_\mathrm{B} =\unit[1.60]{M_\odot}$ PNS with the LS-bulk and CBF-EI EoSs,
\emph{but} with the diffusion coefficients of the GM3 EoS. As expected, we find
out that the LS-bulk timescale is reduced with respect to that of a
self-consistent simulation (i.e., using the LS-bulk diffusion coefficients),
and the CBF-EI timescale is increased with respect to that of a self-consistent
simulation.  Of course, the timescales and the evolutionary profiles found in
this non-consistent manner are not equal to those corresponding to the GM3 EoS,
the differences due to the details of the EoSs. Both the EoS and the neutrino mean free paths influence the PNS
evolution; in fact, each EoS has a different thermal
content and neutrino degeneracy, and different thermodynamical derivatives
that determine how the stellar profiles change while energy and leptons diffuse
through the star.

\subsection{Neutrino luminosity}
\label{ssec:luminosity}

In 1987 a supernova (SN1987a) has been observed in the Large Magellanic
Cloud~\cite{Kunkel+1987}. Together with the electromagnetic signal, 19 neutrinos
were detected by the Cherenkov detectors Kamiokande~II~\cite{Hirata+1987} and
IMB~\cite{Bionta+1987}.  These neutrinos have been observed on a timescale of
ten seconds, and are therefore thought to have been emitted during the PNS
phase. However, they were too few to accurately constrain the emitted neutrino
spectrum and its time dependence (see e.g.~\cite{Lattimer+Yahil.1989}) and to
give unambiguous answers about the proto-neutron star
physics~\cite{Burrows.1988,Lattimer+Yahil.1989,Keil+Janka.1995,Pons+.1999}.
Today, with the current detectors, a SN event such that of 1987 would generate
$\sim10^4$ neutrino detections~\cite{Ikeda+2007}, that would provide valuable
information on the physical processes dominating the
PNS evolution. It is therefore fundamental to determine how the underlying
EoS modifies the PNS neutrino signal.

\begin{figure*}[ht]
\includegraphics[width=\textwidth]{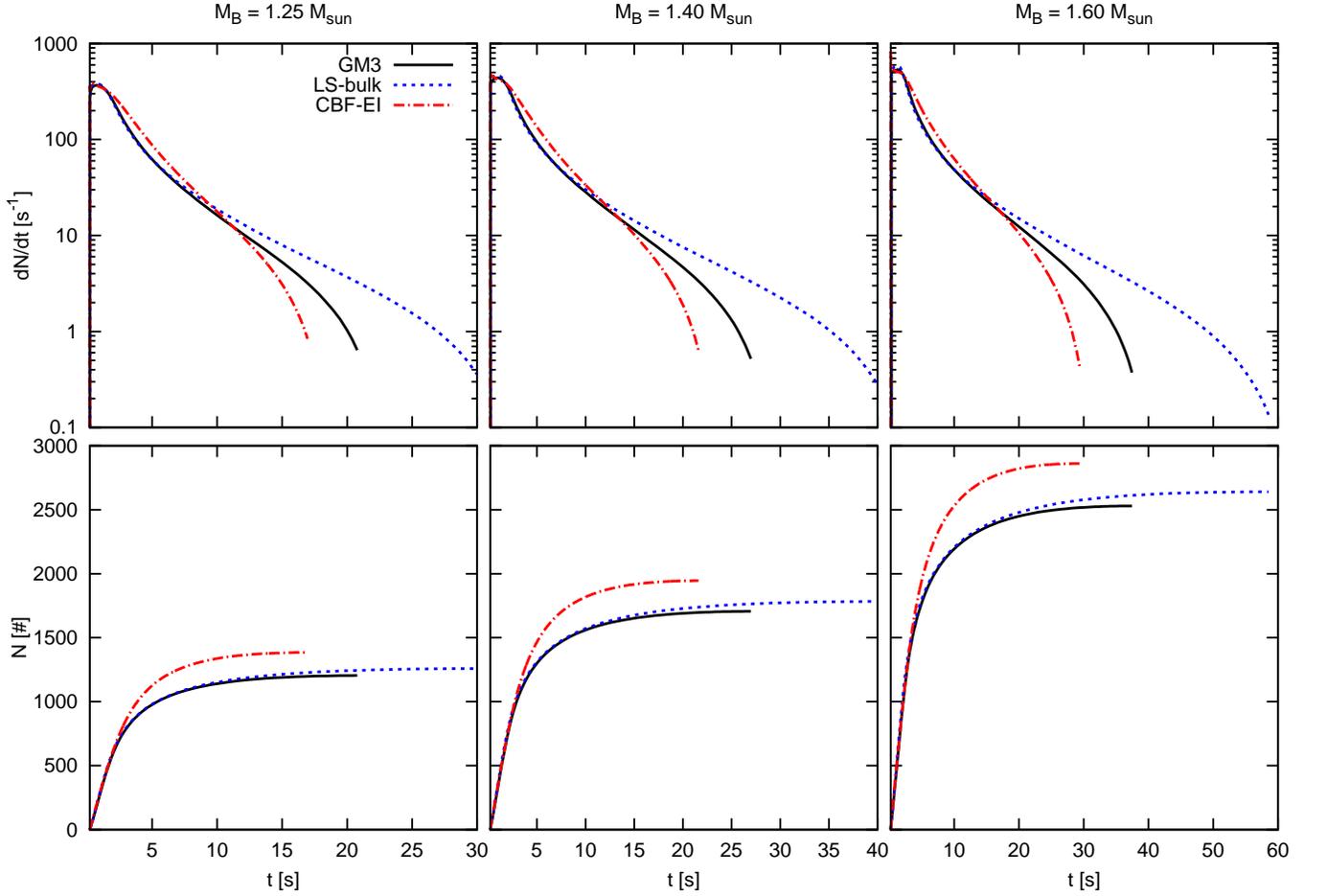}
\caption{Signal in the Super-Kamiokande III Cherenkov detector, for the three
EoSs considered in this paper. In the top panels, electron antineutrino detection
rate; in the bottom panels, electron antineutrino cumulative detection. In the left
plots, we consider a star with $M_\mathrm{B}=\unit[1.25]{M_\odot}$, in the
central plots $M_\mathrm{B}=\unit[1.40]{M_\odot}$, and in the right plots
$M_\mathrm{B}=\unit[1.60]{M_\odot}$. Colors and line styles are as in
Fig.~\ref{fig:L_M_R}.}
\label{fig:detector}
\end{figure*}

Our code has some limitations in reconstructing the emitted spectrum; besides
the spherical symmetry it assumes: (i) beta
equilibrium, (ii) a Fermi distribution for all neutrino species, and (iii) a
vanishing chemical potential for the muon and tauon neutrinos everywhere in the
star.  The assumptions (i) and (ii) are reasonable in the interior of the star,
and lose accuracy near the stellar border, where the diffusion approximation
breaks down and in practice the fluxes are always flux-limited. To obtain a
precise description of the neutrino emitted spectrum, one has to employ
multi-flavour multi-group evolutionary codes (see e.g.~\cite{Roberts.2012}),
that possibly also account for neutrino leakage near the stellar border.  This
is outside the aims of our work; however our approximations are reasonable as
far as one is interested in total quantities, in particular the total neutrino
luminosity $L_\nu$ (Fig.~\ref{fig:L_M_R}), which is equal to minus the
gravitational mass variation rate,
\begin{equation}
\label{eq:Lnu_Mdot}
L_\nu=\mathrm e^{2\phi(R)}4\pi R^2H_\nu(R)=-\frac{\mathrm dM}{\mathrm dt},
\end{equation}
where $H_\nu(R)$ is the neutrino energy luminosity at the stellar border.

We determine the formula to estimate the signal in terrestrial detectors
following~\cite{Burrows.1988} and applying a slight modification introduced
by~\cite{Pons+.1999}, and we specify our results for the Super-Kamiokande III
detector~\cite{Hosaka+2006,Ikeda+2007}.  The main reaction that occurs in a
water detector like Super-Kamiokande is the electron antineutrino absorption on
protons, $\bar\nu_e + p\rightarrow n+e^+$ (Eq.~(1) of~\cite{Ikeda+2007}). The
number flux of antineutrinos arriving at the detector is given by
\begin{align}
\frac{\mathrm d\mathcal N}{\mathrm dt} ={}& \frac{\tilde\sigma_0 \tilde n_p
\mathcal M}{4\pi D^2}\mathrm e^{\phi_\nu} T_\nu L_{\bar \nu_e}\frac{G_W(\mathrm
e^{\phi_\nu}T_\nu,E_\mathrm{th})}{7\pi^4/120},\\
G_W={}&\!\int_{E_\mathrm{th}/T}^\infty\!\!\!  \frac{x^2\left(x-\frac\Delta
T\right)\sqrt{\left(x-\frac \Delta T\right)^2-\left(\frac {m_e}
T\right)^2}}{1+\mathrm e^{x}} W(xT)\mathrm dx,
\end{align}
where $\tilde n_p\simeq\unit[6.7\times10^{31}]{kton^{-1}}$ is the number of free
protons (i.e., hydrogen atoms) per \emph{unit water mass} of the detector,
$\tilde \sigma_0 = \unit[0.941\times10^{-43}]{cm^2MeV^{-2}}$, $\mathcal M$ is
the water mass of the detector, $D$ is the SN distance from the detector, $G_W$
is a modified and truncated Fermi integral, $E_\mathrm{th}$ is the incoming
neutrino energy threshold (to cut off the low-energy neutrino background that is
a noise for high-energy SN and PNS neutrinos,~\cite{Ikeda+2007}), $\Delta_m$ is
the neutron-proton mass difference, $m_e$ is the electron mass, and $W(E)$ is
the efficiency of the detector at incoming neutrino energy $E\equiv xT$.
$\mathrm e^{\phi_\nu}$, $T_\nu$, and $\mu_{\bar\nu_e}$ are the redshift,
temperature, and antineutrino chemical potential at the neutrinosphere, that is
the sphere inside the PNS at whose radius $R_\nu$ neutrinos decouple from matter
(therefore, $\mathrm e^{\phi_\nu} T_\nu$ and $\mathrm
e^{\phi_\nu}\mu_{\bar\nu_e}$ are the temperature and the chemical potential at
the neutrinosphere, seen by an observer at infinity).

We take Super-Kamiokande III as reference detector, and therefore $\mathcal
M\simeq\unit[22.5]{ktons}$~\cite{Ikeda+2007}, $E_\mathrm{th}=\unit[7.5]{MeV}$,
and $W$ is reported in Fig.~3 of~\cite{Hosaka+2006} and is one for
$E>E_\mathrm{th}$. We consider a galactic PNS, $D=\unit[10]{kpc}$, and
assume that the neutrinosphere is at the radius at which the (total, of all flavours)
neutrino energy flux becomes one third of the (total, of all flavours) neutrino
energy density, $H_\nu/\epsilon_\nu=1/3$. Finally, we take the electron
antineutrino energy to be one sixth of the total, $L_{\bar\nu_e}=L_\nu/6$, since
(i) at the neutrinosphere all neutrino type chemical potentials are very small
and (ii) we do not account for neutrino oscillations (which would enhance the
flux by about $10\%$~\cite{Ikeda+2007}).

The neutrino signal rate and total signal for the three EoSs are shown in
Fig.~\ref{fig:detector}.  Since the binding energies of the cold neutron star of
the three EoSs we consider are very similar (see Fig.~\ref{fig:L_M_R}), the total
energies emitted by neutrinos during the PNS evolution are very similar too. On the
other hand, the rate of antineutrino emission and the temperature at the
neutrinosphere varies according to the underlying EoS. Therefore, there is an
EoS signature on the cumulative antineutrino detection. The signal of the CBF-EI
PNS is noticeably larger than the other EoSs, even though its gravitational
binding energy at the end of the evolution is between those of the LS-bulk and
GM3 EoSs  (Fig.~\ref{fig:L_M_R}). This is due to the fact that the higher
temperatures of the CBF-EI EoS cause a smoother antineutrino distribution
function at the neutrinosphere, and hence more antineutrinos have an energy
greater than the threshold $E_\mathrm{th}$ at the detector.

The different evolutionary timescales for the three EoSs and stellar masses
correspond to different signal timescales, that may easily be inferred from the
antineutrino detection rate.  The antineutrino detection rates 
for the three EoSs and the three stellar masses are qualitatively
very similar.  During the
first ten seconds the LS-bulk and GM3 stars have very similar detection rates;
at later times, the detection rates become different, because the LS-bulk star
has a longer evolution than the GM3 star. The CBF-EI star, instead, has the
peculiarity of maintaining a higher antineutrino emission rate during the
Joule-heating phase (approximately, during the first ten seconds), which is due
to the faster deleptonization that we have already discussed in
Sec.~\ref{ssec:evolution} and to higher temperatures.

\section{Gravitational waves from quasi-normal modes}
\label{sec:GW}

A supernova explosion is a highly energetic event and the  PNS which is formed
as a remnant is expected to oscillate wildly. The relativistic theory of stellar
perturbations~\cite{Thorne+Campolattaro.1967,chandrasekhar1991non} predicts
the existence of stellar oscillation modes, the so-called quasi-normal modes
(QNMs), through which the star loses energy emitting gravitational waves (GWs).
To find the frequencies of these modes,
in the case of a spherical, non rotating star, Einstein's equations are perturbed about the 
background~\eqref{eq:metric}, and the perturbed functions are expanded
in spherical harmonics and Fourier-transformed.  
Thus, the spacetime metric describing the perturbed spacetime
can be written as\footnote{We  use the 
gauge adopted in \cite{Thorne+Campolattaro.1967}.}
\begin{multline}
\label{eq:metric_oscillations}
\mathrm ds^2=-\mathrm e^{2\phi} (1+r^l H_0Y_{l m}\mathrm e^{\imath\omega
t+\imath m\varphi})\mathrm dt^2\\
-2\imath\omega r^{l+1} H_1Y_{l m}\mathrm e^{\imath\omega t+\imath
m\varphi}\mathrm dt\mathrm dr\\
+ \mathrm e^{2\lambda} (1-r^lH_0Y_{l m}\mathrm e^{\imath\omega t+\imath
m\varphi})\mathrm dr^2\\
+ r^2(1-r^lKY_{l m}\mathrm e^{\imath \omega t+\imath m\varphi})\mathrm
d\Omega\,,
\end{multline}
where $Y_{lm}(\vartheta,\varphi)$ are the scalar spherical harmonics and
$H_0(r,\omega)$, $H_1(r,\omega)$, $K(r,\omega)$ describe the polar metric
perturbations. A fluid element in a point $x^\mu$ is displaced by the
perturbation in the new position $x'^\mu=x^\mu+\xi^\mu$, where the displacement
vector $\xi^\mu$ can be written as
\begin{align}
\label{eq:displacement_r}
\xi^t={}&0,\\
\xi^r={}&r^{l-1}\mathrm e^{-\lambda}W(r,\omega) Y_{l m}(\theta,\varphi) \mathrm
e^{\imath\omega t+\imath m\varphi},\\
\label{eq:displacement_theta}
\xi^\theta={}&-r^{l-2} V(r,\omega)\partial_\theta Y_{l m}(\theta,\varphi)
\mathrm e^{\imath\omega t+\imath m\varphi},\\
\label{eq:displacement_phi}
\xi^\varphi={}&-\frac{r^{l-2}{V(r,\omega)}}{\sin^2\theta}\partial_\varphi
Y_{lm}(\theta,\varphi) \mathrm e^{\imath\omega t+\imath m\varphi}\,.
\end{align}
The perturbations of the energy density and pressure of the fluid composing the
star are expanded in the same way.  Due to the decomposition in spherical
harmonics and to the Fourier expansion, the linearized Einstein+hydro equations
do separate, and  are reduced to a set of coupled, linear ordinary differential equations for the
radial part of the perturbed fluid and of the metric functions.  

A QNM is defined as a solution of the perturbed equations which is regular at
the center, continuous at the stellar surface, and which behaves as a purely
outgoing wave at radial infinity.  
The set of discrete values of the complex
frequency $\omega=2\pi\nu+\imath/\tau$ for which these conditions are
satisfied are the QNM eigenfrequencies:
the real part is the pulsation frequency $\nu$, 
the imaginary part is the inverse of  the damping time $\tau$. 

The QNMs are classified according to the nature of the restoring force which
prevails in bringing back the perturbed fluid element to the equilibrium position.
For the $p_n$-modes, or ``pressure modes'', ($n=1,2,\dots$) the main restoring force is due to
pressure; for the  $g_n$ modes ($n=1,2,\dots$), or ``gravity modes'', the
main restoring force is buoyancy.
The order $n$ of the mode corresponds to the
number of nodes of the radial eigenfunction of the  displacement vector. The
$f$-mode, i.e., the fundamental mode of the star, describes the global pulsation
motion of the fluid, and has no radial nodes. In a cold neutron star, typical values for
the QNM frequencies and damping times are $\nu_f\simeq\unit[1.5-2.5]{kHz}$,
$\tau_f\simeq\unit[0.1]{s}$, $\nu_{p_1}\simeq\unit[5-10]{kHz}$, and
$\tau_{p_1}=\unit[1-10]{s}$. 
The $g$-modes are due to the
presence of thermal and/or composition gradients; in absence of composition
gradients, all $g$-modes of a cold neutron star degenerate to zero
frequency. Conversely, they are present in a PNS
\cite{Ferrari+Miniutti+Pons.2003,Burgio+2011}, as we shall show below.

To determine the quasi-normal mode frequencies at a given time $t$ of the stellar evolution, we have first evolved the
PNS, finding the profiles of the pressure $P(r,t)$, the energy density $\epsilon(r,t)$, the baryon number density
$n_\mathrm{B}(r,t)$, and the sound speed, $c_s(r,t)$, for the three EoSs and the different values of the baryonic mass
we consider in this paper.  Then we have determined the ``effective barotropic EoS'' by inverting the pressure-radius
profile, thus finding $r= r(P,t)$ and then $\epsilon^{eff}(P;t)=\epsilon(r(P,t),t)$ and $c^{eff}_s(P;t)=c_s(r(P,t),t)$.
Using these expressions, we have solved the equations of stellar
perturbations (we used the formulation of~\cite{Detweiler+Lindblom.1985}), to find the frequencies and damping times of
the first $p$- and $g$-modes and of the fundamental mode.

\subsection{Results of the numerical evolution}
\begin{figure}[ht]
\centering
\includegraphics[width=\columnwidth]{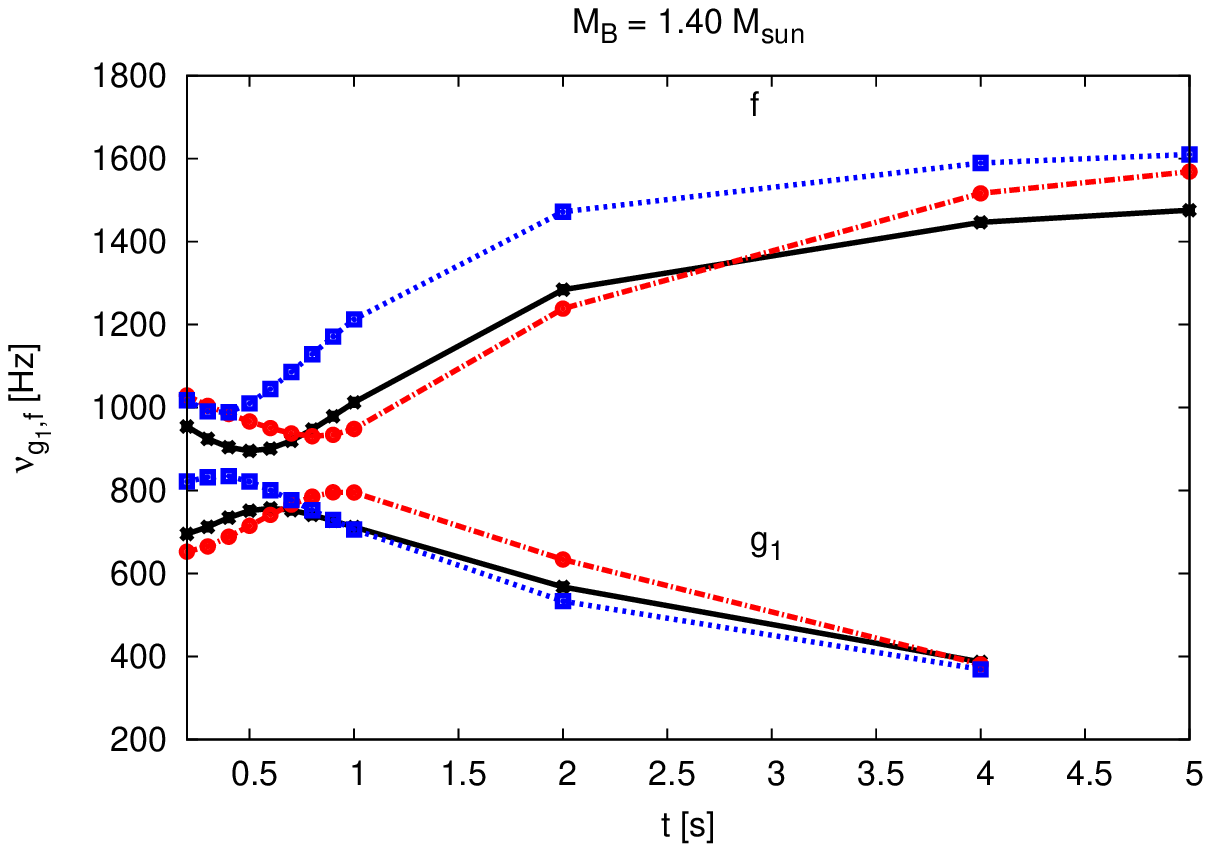}
\includegraphics[width=\columnwidth]{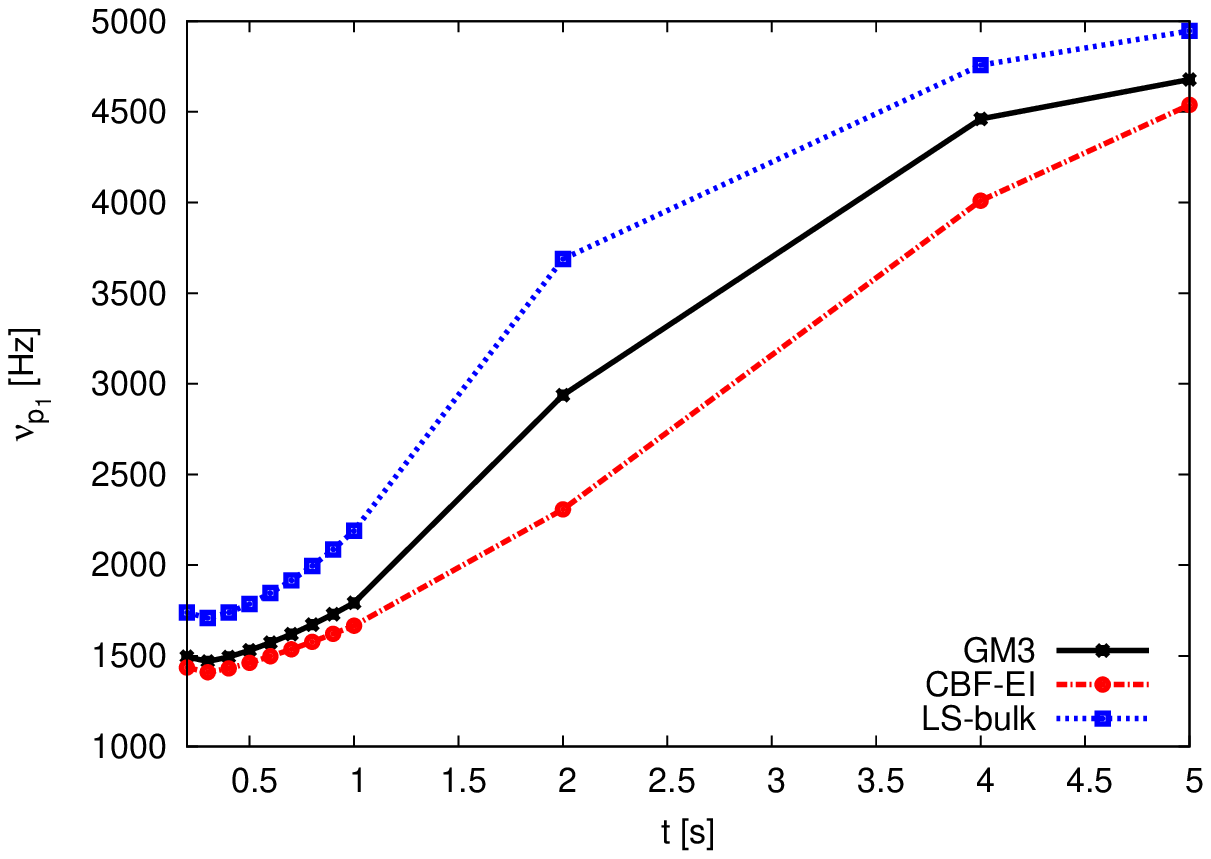}
\includegraphics[width=\columnwidth]{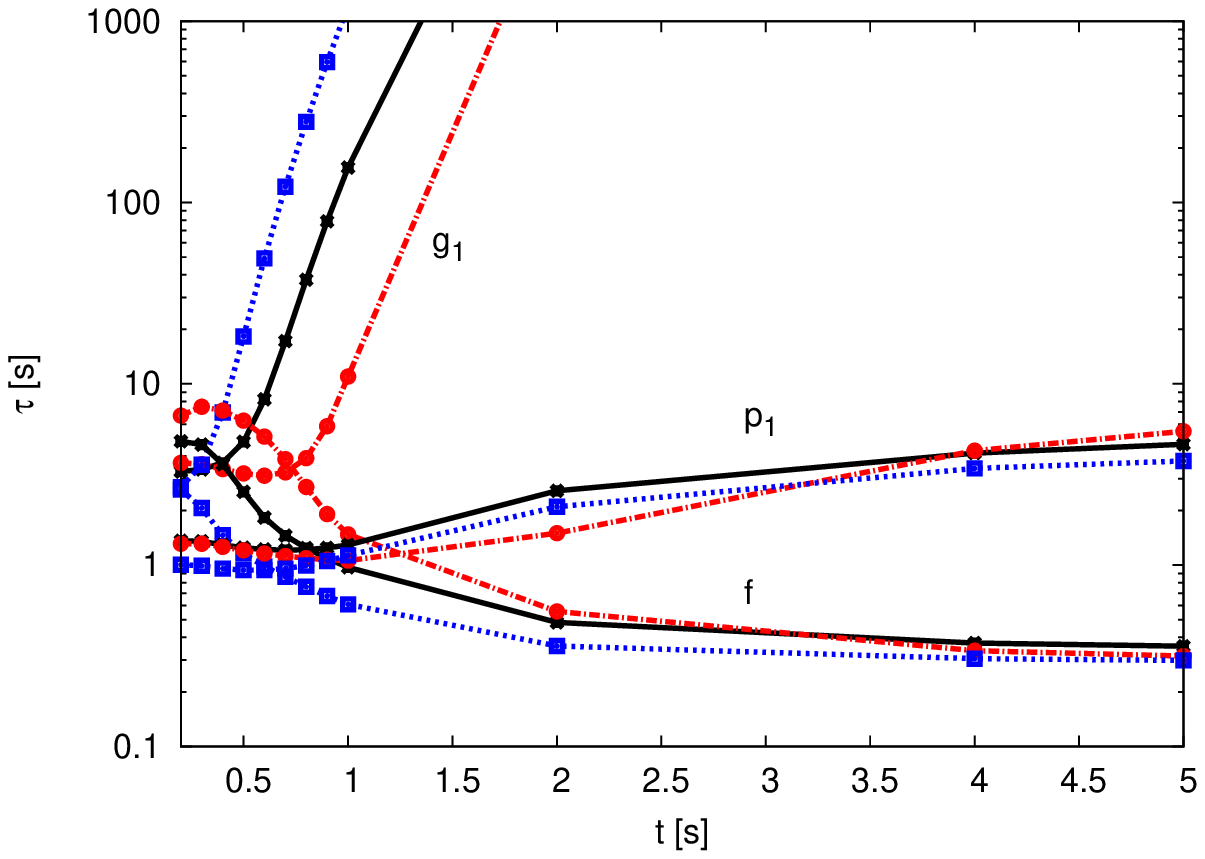}
\caption{Time dependence of the PNS quasi-normal mode frequencies and damping
times for the three EoSs and for $M_\mathrm{B}=\unit[1.40]{M_\odot}$.}
\label{fig:QNM_1.40}
\end{figure}
We have evolved three stellar models with baryon masses
(1.25, 1.40, and 1.60 $\unit{M_\odot}$) and the  EoSs LS-bulk, CBF-EI and
GM3, which was used in \cite{Ferrari+Miniutti+Pons.2003}.  
For this EoS, the QNM frequencies we compute for the $\unit[1.60]{M_\odot}$ star
agree with those of ``model A''
of~\cite{Ferrari+Miniutti+Pons.2003} within a few percent.
We think that the small differences between our results
and those of~\cite{Ferrari+Miniutti+Pons.2003} are due to differences in the
initial profiles and in the details of the treatment of the diffusion processes.
The numerical values of the $f$-, $g_1$- and $p_1$- QNM frequencies and damping
times are tabulated in Appendix~\ref{app:tables}.

In Fig.~\ref{fig:QNM_1.40} we show, as an example,  how the QNM frequencies and
damping times change during the first 5 seconds of the PNS life.
The plots are given for the three EoSs we consider, and for a star with baryonic
mass $M_\mathrm{B}=\unit[1.40]{M_\odot}$ as an example. 

In the upper panel we show the frequency of the $g_1$- and of the $f$- modes, in
the mid panel the frequency of the  mode $p_1$, and in the lower panel
the damping time of the three modes.  
From the upper panel of Fig.~\ref{fig:QNM_1.40} we see that during the first
second, $\nu_{g_1}$ approaches $\nu_f$, but they never cross.  At later times,
$\nu_{g_1}$ increases, reaches a maximum and then decreases, whereas $\nu_f$ does the
opposite: it reaches a minimum slightly before $\nu_{g_1}$ reaches its maximum,
and then increases toward the asymptotic value of the corresponding cold neutron
star. This behaviour is a general feature of the three EoS;
however, the minimum (maximum)
of $\nu_f$ ($\nu_{g_1}$) occurs at different times for different EoSs.
In addition frequencies belonging to different EoSs differ, at each time, as much as
$\sim 100$--$\unit[200]{Hz}$.
$\nu_{p_1}$ also has a minimum (which was not found in~\cite{Ferrari+Miniutti+Pons.2003}),
at earlier times with respect to $\nu_f$ and $\nu_{g_1}$.

It may be noted that our results are qualitatively different from those
of~\cite{Burgio+2011} and~\cite{Sotani:2016uwn}, where the QNMs show a monotonic
increase of the $f$- and $p$-modes, and a monotonic decrease of the $g$-mode.
We think that this is due to the fact that a consistent evolution of the
PNS is  crucial to describe the behaviour of the QNMs.

The time dependence of the QNM frequencies described above would produce differences in the
gravitational waveforms emitted by the PNS which, if detected, 
would provide valuable information on the underlying EoS.   
The waveform emitted by a star oscillating 
in a QNM with frequency $\nu$ and
damping time $\tau$ can be written as
$h(t)=h_0\mathrm e^{-(t-t_0)/\tau}\sin[2\pi\nu(t-t_0)]$,
where $h_0$ is the initial amplitude and $t_0$ some initial time.  Since
the mode energy is proportional to the square of the wavefunction,
$E_\mathrm{QNM}\propto\mathrm e^{-2(t-t_0)/\tau}$, and the gravitational wave
(GW) luminosity is $L_\mathrm{GW}=-\dot E_\mathrm{QNM}\simeq
\frac{2E_\mathrm{QNM}}{\tau}.$ Therefore, QNMs with smaller damping times are
more effective in extracting energy from the PNS in the form of GWs.  In a cold
star $\tau_f < \tau_p$, and this means that the energy will be radiated mainly
at the frequency of the  fundamental mode.  However, during the first second of
the PNS life the situation is quite different; the lower panel of
Fig.~\ref{fig:QNM_1.40}  shows that the $p_1$-mode has a damping time
$\tau_{p_1}\simeq\unit[1]{s}$, \emph{smaller} than that of the $f$- and of the
$g_{1}$-modes, and it can be \emph{more effective} in radiating energy than the
fundamental and the first $g$-mode. After the first second, the fundamental mode
becomes the more efficient GW emitter.

It should be stressed that the mechanical energy of a newly born PNS is
dissipated in gravitational waves only in part. GWs compete with other
dissipative mechanisms associated to neutrino diffusion; therefore,
gravitational waves will be emitted by a PNS  only if $\tau_{GW}$ is smaller
than the dissipation timescales typical of neutrino diffusion.  These have been
estimated to be of the order of $\tau_{\nu}\sim 10-20$ s (see
~\cite{Ferrari+Miniutti+Pons.2003} for a discussion on this issue and references
therein).  From the lower panel of Fig.~\ref{fig:QNM_1.40} we see that the
damping times  of the $f$- and $p_1$- modes are always smaller than
$\tau_{\nu}$, whereas $\tau_{g_1}$ becomes larger  than $\tau_{\nu}$ after the
first few tenths of seconds.  Thus, if the PNS has a significant amount of mechanical
energy to release, we can reasonably expect that a part of it will be released
in gravitational waves.  

Recent 3-D simulations of the early explosion phase of
core-collapse supernovae and of the following accretion
phase~\cite{Andresen:2016pdt,Kuroda:2016bjd}
show that other phenomena than stellar oscillations may
contribute to gravitational wave emission; for instance,  standing accretion shock
instability and convection, which are shown to be associated to stochastic oscillations,
and to unstable $g$-modes, different from the stable $g$-modes considered in this paper.
For a review see also~\cite{Kotake:2011yv}.

\subsection{A fit of the fundamental and first p-mode periods}
\label{ssec:fit_GW}

\begin{figure}
\includegraphics[width=\columnwidth]{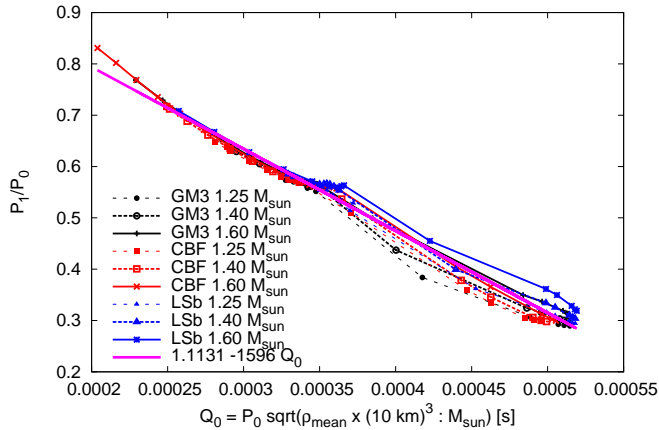}
\caption{Ratio between the first p-mode period and the fundamental period as a
function of the quantity $Q_0$, for a PNS in the configurations studied in this
paper. The points at the top-left correspond to late evolutionary times, whereas
the initial configurations are in the bottom-right.  See text for details.}
\label{fig:P1oP0vsQ0}
\end{figure}

In non-relativistic variable stars (as Chepheids), the ratio of the periods
$P_1/P_0=\nu_f/\nu_{p_1}$ of the first overtone (that corresponds in the
language of stellar oscillations in GR to the first p-mode) and of the
fundamental mode is a function of the quantity $Q_0=P_0\sqrt{\bar
\rho/\rho_\odot}$, where $P_0=\nu_f^{-1}$ and
$\rho_\odot=\unit[2.97\times10^{-18}]{M_\odot/km^3}$ is the mean Sun density
(see e.g.~\cite{Christy.1966}). We have fitted the ratio
$P_1/P_0\equiv\nu_f/\nu_{p_1}$ with a linear dependence on
$Q_0\propto\sqrt{\bar\rho}/\nu_f$, obtaining
\begin{equation}
\label{eq:fitGW}
\frac{P_1}{P_0}=1.1131(\pm0.0066) - 1596(\pm17)\frac{P_0}{\unit[1]{s}}\sqrt{\bar
\rho \frac{\unit[10^3]{km}}{\unit[1]{M_\odot}}}.
\end{equation}
The result of the fit is shown in Fig.~\ref{fig:P1oP0vsQ0}; the corresponding
reduced chi-square is rather low: $\tilde\chi=3.2\times10^{-4}$. This is an
indication that, even in PNSs, the ratio $P_1/P_0$ could be an universal
property, independent of the masses and EoSs of the PNS.
\section{Conclusions}
\label{sec:conclusions}

In this paper we have studied the evolution and the gravitational wave emission of a proto-neutron star in the
Kelvin-Helmholtz phase, that is the period of the neutron star life subsequent to the supernova explosion, until the
star becomes transparent to neutrinos.  To perform such a study, we have written a new general relativistic,
one-dimensional, energy-averaged, and flux-limited PNS evolutionary code which evolves a general EoS consistently.  In
particular, we have considered three nucleonic EoSs and three stellar masses, and we have determined the neutrino cross
sections self-consistently with the corresponding EoS. The EoSs considered are all nucleonic (without hyperons) and are
obtained (i) by the extrapolation from the measured nuclear properties (the LS-bulk EoS~\cite{Lattimer+Swesty.1991})
(ii) by the nuclear relativistic mean-field theory (the GM3 EoS~\cite{Glendenning+Moszkowski.1991}) (iii) by the nuclear
non-relativistic many-body theory (the CBF-EI EoS~\cite{Benhar+Lovato.2017, Lovato+Benhar.prep}).  We have
determined the frequencies of the quasi-normal oscillation modes for the different EoSs and stellar masses using the
general relativistic stellar perturbation theory.

The main improvements with respect to previous works introduced by our
study are the following.
\begin{itemize}
\item We have developed and tested a new fitting formula for the
interacting part of the baryon free energy (i.e., neutrons plus
protons), which is valid for high density matter, finite temperature,
and arbitrary proton fractions.  We used this fitting formula to
derive the other thermodynamical quantities.  This formula is suitable
to be used in evolutionary codes.
\item We have computed the neutrino cross sections for the many-body
theory EoS of~\cite{Benhar+Lovato.2017, Lovato+Benhar.prep}.  They have been computed at
the mean-field level~\cite{Reddy+Prakash+Lattimer.1998}, that is, the
interaction between baryons has been accounted for modifying the
baryon energy spectra by means of density-, temperature-, and
composition-dependent effective masses and single-particle energies.
\item We used these neutrino cross sections to evolve the PNS with the
many-body EoS in a consistent way.  To our knowledge, this is the
first time that a PNS with a many-body EoS has been evolved with
consistently determined neutrino opacities. From this  evolution, we
have determined the stellar quasi-normal modes.
\end{itemize}

Our main results are the following.
\begin{itemize}
\item The PNS evolution depends on the adopted EoS. In particular, for the many-body EoS CBF-EI the PNS cooling is
  faster than that with the mean-field EoS GM3, which in turn is faster than that with the extrapolated EoS LS-bulk.
In the extrapolated EoS
LS-bulk the effective baryon masses have been assumed to be equal to the bare ones,
and the result is that this EoS is ``less interacting'' than the others in
the computation of the neutrino cross sections.
\item The deleptonization of the PNS with the CBF-EI EoS is almost completed at the end of the Joule-heating phase
  (similarly to what was found in the first PNS numerical studies by~\cite{Burrows+Lattimer.1986}
  and~\cite{Keil+Janka.1995}), whereas the deleptonization for the GM3 and LS-bulk EoSs proceeds during the cooling
  phase (as found in~\cite{Pons+.1999}).  Pons {\it et al.}~\cite{Pons+.1999} explained this difference with the
  over-simplifications in the treatment of the neutrino opacities in~\cite{Burrows+Lattimer.1986,Keil+Janka.1995}.
  However, we compute the neutrino cross sections for the CBF-EI and the other EoSs with the same procedure
  of~\cite{Pons+.1999}.  Therefore, the faster deleptonization is a feature also due to the EoS properties and not only
  to the treatment of neutrino opacities.
\item The total number of electron antineutrinos detected depends on the
gravitational binding energy but is not completely determined by it. In
particular, the CBF-EI EoS has more antineutrinos detected than the LS-bulk EoS,
even though its binding energy is smaller. This is due to the fact that the PNS with CBF-EI EoS
has higher temperatures than those with the other EoSs, hence the electron antineutrino
distribution function at the neutrinosphere is smoother and more antineutrinos
have energies larger than the detector energy threshold.  This
result remarks the importance of an accurate modeling of the PNS evolution in
order to extract information on the PNS physics from the neutrino signal.
\item 
We show that during the first second, the frequencies at which the PNS
oscillates emitting  gravitational waves have a non monotonic behaviour.
The fundamental mode frequency decreases, reaches a minimum and then increases 
toward the value corresponding to the  cold neutron star which forms at
the end of the evolution. The frequency of the first $g$ mode increases, reaches
a maximum and then decreases to the asymptotic zero limit, that of the mode $p_1$ has  
a less pronounced minimum at earlier times with respect to the $f$ mode. We show
that this behaviour, already noted in~\cite{Ferrari+Miniutti+Pons.2003} for the EoS GM3,
is a generic feature when the PNS evolution is consistently described, and that
the timescale depends on the EoS. Indeed the time needed to reach the minimum
(maximum) for the $f$- ($g_1$-) mode can differ by as much as half a second 
for the EoS we consider.

During the first second, the damping time of all modes  is shorter than the
neutrino diffusive timescale ($\sim \unit[10]{s}$);  therefore gravitational
wave emission may be competitive in subtracting energy from the star.  This
remains true at later time only for the fundamental mode and for the first $p$
mode. However, the damping time of the $f$ mode is much shorter, thus we should
expect that after the first few seconds gravitational waves will be emitted
mainly at the corresponding frequency.

\item The QNM frequencies depend not only on the EoS, but also on the stellar baryon mass. In
particular, we find that for a lower mass, at the beginning the $p_1$-mode has 
higher frequency; for instance, for the $\unit[1.25]{M_\odot}$ star it approaches
2 kHz.
\item We have found a relation between the fundamental and first p-mode
frequencies and the mean stellar density [Eq.~\eqref{eq:fitGW}] which is valid
during all PNS phases for the cases considered in this paper. This may be an
universal property of PNSs, independent of the mass and the EoS.
\end{itemize}

This paper may be improved in several directions. About the microphysics,
improvements may be done to the neutrino cross section treatment, for example
including the effects of collective excitations~\cite{Lovato+Losa+Benhar.2013,
Lovato+2014} and the weak magnetism correction~\cite{Roberts+Reddy.2016}.
Consistently computed neutrino cross sections in the many-body theory for
finite temperature and high density matter would be welcome too. About the EoS,
it would be interesting to include more physical ingredients, like hyperons and
the presence of a crust (alpha particles and a lattice). However, we do not
expect dramatic changes for the inclusion of a crust since we have checked that
alpha particles do form only near the stellar surface and towards the end of
our simulations (Appendix~\ref{ssec:NSE}). About the PNS evolution, it would be
interesting to abandon the request of beta equilibrium (even though, we have
checked that beta equilibrium is almost respected in most of the star, apart
for a region near the stellar layer at the beginning of the simulation, see
Appendix~\ref{ssec:beta-equilibrium}
and~\cite{Burrows+Lattimer.1986,Pons+.1999}) and to allow for the presence of
muons and tauons, accounting for the transport of their relative lepton
numbers.
A major improvement to our work would be the inclusion of
accretion~\cite{Burrows.1988, Takiwaki+Kotake+Suwa.2014,
Melson+Janka+Marek.2015, Muller.2015} and convection~\cite{Roberts+2012}, which could
have an important effect on the evolution.  Finally, we are using as initial
configurations the final profile obtained from an old
simulation~\cite{Wilson+Mayle.1989}, conveniently rescaling it with the total
stellar mass. This brings a significant amount of uncertainties; to increase
the accuracy of the evolution it would be important to consistently use the
final profiles of more modern core-collapse simulations.

\begin{acknowledgments}
This work has been partially supported by “NewCompStar,” COST Action MP1304,
by the H2020-MSCA-RISE-2015 Grant No. StronGrHEP-690904,
by INFN (through grant TEONGRAV),
and supported in part by the U.S. Department of Energy, Office of Science, Office of Nuclear Physics, under Contract No. DE-AC02-06CH11357 (A. L.).
J.A.P. acknowledges support by the Spanish MINECO grant AYA2015-66899-C2-2-P.

\end{acknowledgments}
\appendix
\section{Fitting procedure}
\label{app:fit}

The fit has been performed using a set of points on an evenly spaced Cartesian
$11\times50\times12$ grid in $(Y_p;T;n_\mathrm{B})$, from
$(0;\unit[1]{MeV};\unit[0.04]{fm^{-3}})$ to
$(0.5;\unit[50]{MeV};\unit[0.48]{fm^{-3}})$, with steps of
$(0.05;\unit[1]{MeV};\unit[0.04]{fm^{-3}})$.
The fit is strictly valid for $n_\mathrm{B}\in(0.04;0.48)\,\unit{fm^{-3}}$ and
$T\in(1;50)\,\unit{MeV}$, but its analytic form is suitable
to be used also for $n_\mathrm{B}<\unit[0.04]{fm^{-3}}$ and $T<\unit[1]{MeV}$, see Sec.~\ref{ssec:fit}. First, we have
fitted \emph{only} the interacting free energy $f^I_\mathrm{B}$, computing the
root mean square $\sigma_f$.  We have done the same for the interacting entropy and
pressure, obtaining the root mean squares $\sigma_s$ and $\sigma_P$.  Then, we have
\emph{simultaneously} fitted the interacting free energy, entropy, and pressure, giving to each fitting point $p_i$
an uniform error $\sigma_{i=\{f;s;P\}}$ that depends on which quantity that
point is describing (the free energy, the entropy, or the pressure).  The
result of the fit of the GM3 and CBF-EI EoS, is shown in Tab.~\ref{tab:f_fit}. We have tried to include
in the fit also the second order derivatives, $\partial^2 f_\mathrm{B}/\partial
T^2$, $\partial^2 f_\mathrm{B}/\partial n_\mathrm{B}^2$, and $\partial^2
f_\mathrm{B}/\partial T\partial n_\mathrm{B}$, but the resulting fit did not improve its accuracy.

We have checked that in the range we consider, the fits for the
GM3 and the CBF-EI EoSs satisfy the thermodynamic stability conditions
(Eqs.~(13) and (14) of~\cite{Swesty.1996})
\begin{align}
\left.\frac{\partial s_\mathrm{B}}{\partial T}\right|_{n_\mathrm{B}}>{}&0,\\
\left.\frac{\partial P_\mathrm{B}}{\partial n}\right|_T>{}&0.
\end{align}
\begin{table}
\caption{Interacting baryon free energy per baryon fitting parameters, Eqs.~\eqref{eq:ffit} and
\eqref{eq:fSNM_PNM}. In the first column, we report the fitting coefficient for SNM and PNM, in
the second and third columns we report the results of the fit for the GM3 and
CBF-EI EoSs, in the fourth and last column, we report the polynomial that is
multiplied by that coefficient in the fitting formula.  In the last two rows we
report the number of points used in the fit and the reduced chi-squared. 
Energies are in MeV and lengths in fm.}
\label{tab:f_fit}
\centering
\begin{tabular}{cccl}
coeff. & GM3 & CBF-EI & polynomial \\
\hline
$a_{1,\mathrm{SNM}}$ & $-402.401    $ & $-284.592   $ & $4Y_p(1-Y_p)n_\mathrm{B}$      \\
$a_{2,\mathrm{SNM}}$ & $1290.54     $ & $676.121    $ & $4Y_p(1-Y_p)n_\mathrm{B}^2$    \\
$a_{3,\mathrm{SNM}}$ & $-1540.52    $ & $-662.847   $ & $4Y_p(1-Y_p)n_\mathrm{B}^3$    \\
$a_{4,\mathrm{SNM}}$ & $903.8       $ & $667.492    $ & $4Y_p(1-Y_p)n_\mathrm{B}^4$    \\
$a_{5,\mathrm{SNM}}$ & $0.0669357   $ & $0.112911   $ & $4Y_p(1-Y_p)n_\mathrm{B} T^2$   \\
$a_{6,\mathrm{SNM}}$ & $-0.000680098$ & $-0.00124098$ & $4Y_p(1-Y_p)n_\mathrm{B} T^3$   \\
$a_{7,\mathrm{SNM}}$ & $-0.0769298  $ & $-0.148538  $ & $4Y_p(1-Y_p)n_\mathrm{B}^2T^2$ \\
$a_{8,\mathrm{SNM}}$ & $0.000915968 $ & $0.00192405 $ & $4Y_p(1-Y_p)n_\mathrm{B}^2T^3$ \\
\hline                                                                 
$a_{1,\mathrm{PNM}}$ & $-274.544    $ & $-121.362    $ & $(1-2Y_p)^2n_\mathrm{B}$      \\
$a_{2,\mathrm{PNM}}$ & $1368.86     $ & $101.948     $ & $(1-2Y_p)^2n_\mathrm{B}^2$    \\
$a_{3,\mathrm{PNM}}$ & $-1609.15    $ & $1079.08     $ & $(1-2Y_p)^2n_\mathrm{B}^3$    \\
$a_{4,\mathrm{PNM}}$ & $916.956     $ & $-924.248    $ & $(1-2Y_p)^2n_\mathrm{B}^4$    \\
$a_{5,\mathrm{PNM}}$ & $0.0464766   $ & $0.0579368   $ & $(1-2Y_p)^2n_\mathrm{B} T^2$   \\
$a_{6,\mathrm{PNM}}$ & $-0.000388966$ & $-0.000495044$ & $(1-2Y_p)^2n_\mathrm{B} T^3$   \\
$a_{7,\mathrm{PNM}}$ & $-0.0572916  $ & $-0.0729861  $ & $(1-2Y_p)^2n_\mathrm{B}^2T^2$ \\
$a_{8,\mathrm{PNM}}$ & $0.00055403  $ & $0.000749914 $ & $(1-2Y_p)^2n_\mathrm{B}^2T^3$ \\
\hline
$N$ & $19782$ & $18686$& \\
$\tilde\chi$ & $4.18$ & $2.05$ & \\
\hline
\end{tabular}
\end{table}

\section{Code checks}
\label{app:code_checks}

In this appendix we show the checks of the accuracy of the code, and we justify
\emph{a posteriori} the assumption of beta equilibrium and the assumption of a
baryon EoS made of an interacting gas with neither alpha particles nor a solid crust.
For simplicity, we show the results of a PNS evolved with the CBF-EI EoS and
with total baryon mass $M_\mathrm{B}=\unit[1.60]{M_\odot}$; the results for
the other EoSs and the other baryon masses are similar.

\subsection{Energy and lepton number conservation}
\label{ssec:conservation}

\begin{figure}
\centering
\includegraphics[width=\columnwidth]{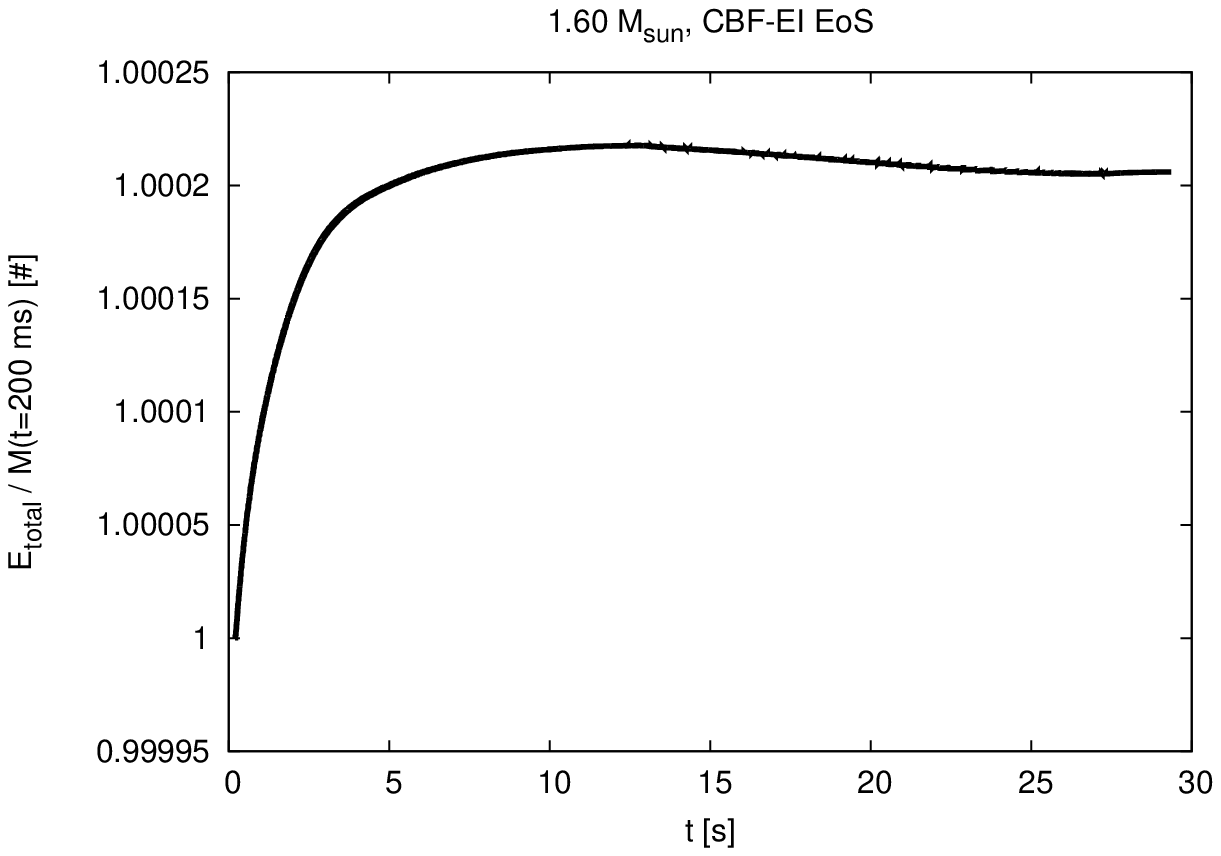}
\includegraphics[width=\columnwidth]{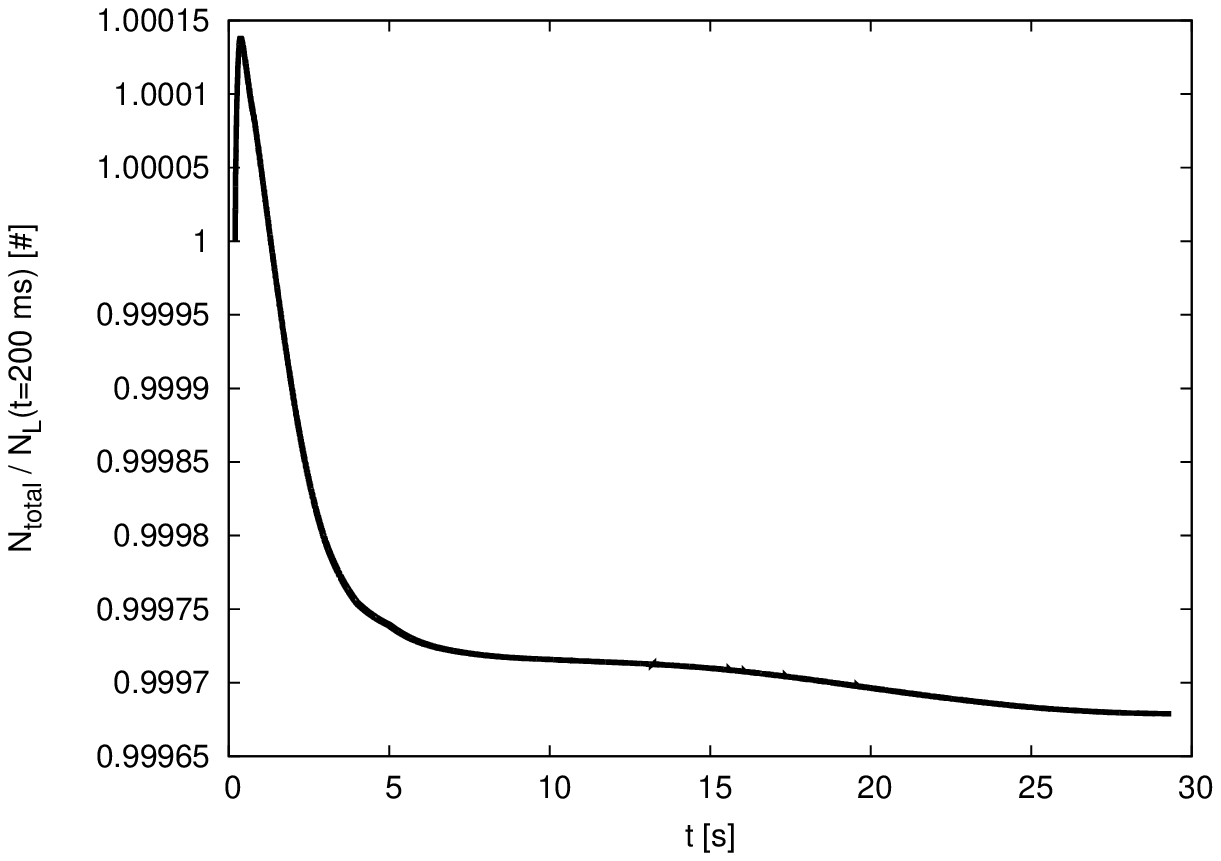}
\caption{Total energy $E_\mathrm{total}$ and total lepton number $N_\mathrm{total}$
conservation for a PNS with the CBF-EI EoS and
$\unit[1.60]{M_\odot}$ baryon mass, normalized with the stellar initial energy and
lepton number (our simulations start at $\unit[200]{ms}$, see Sec.~\ref{ssec:evolution_implementation}). The timestep is changed during the
evolution in such a way that the relative variation in a timestep of the profiles of entropy per baryon and lepton fraction
is approximately equal to $10^{-4}$.
}
\label{fig:conservation}
\end{figure}

\begin{figure}
\centering
\includegraphics[width=\columnwidth]{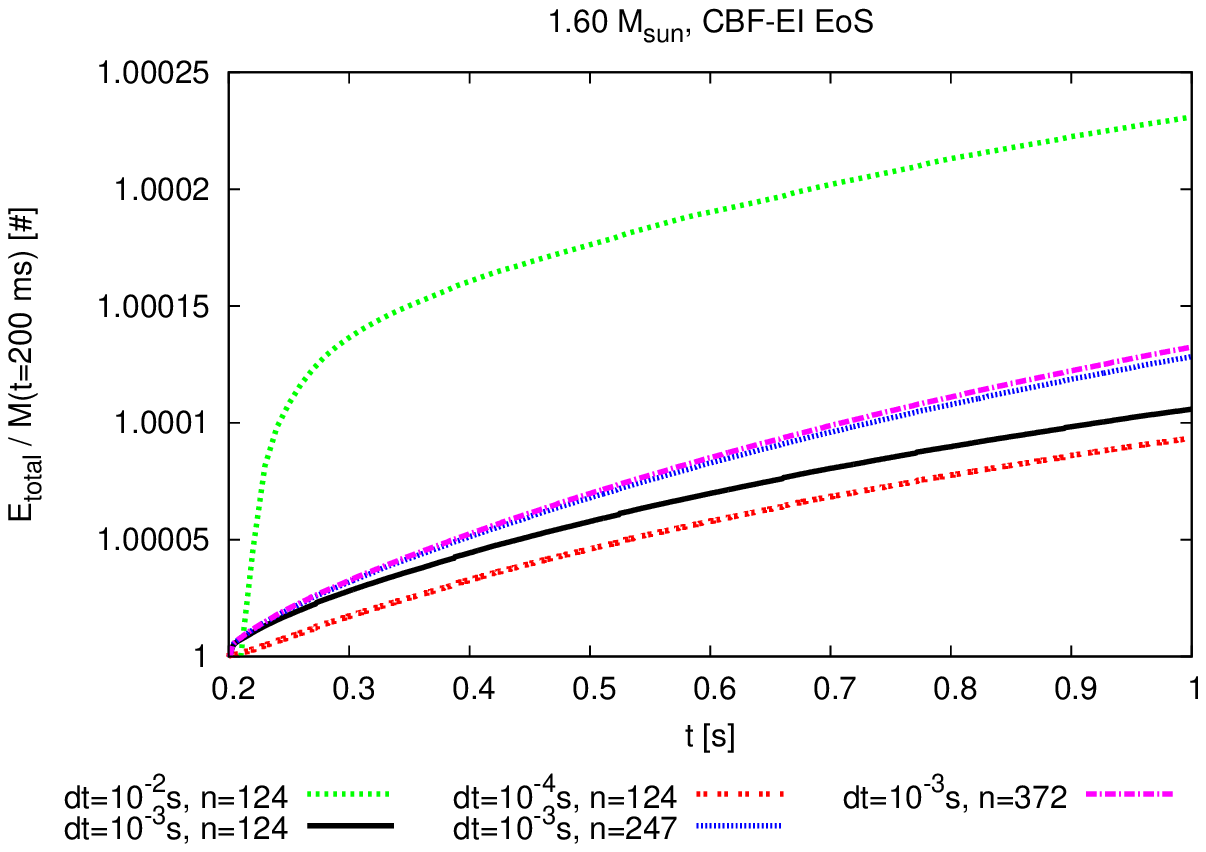}
\includegraphics[width=\columnwidth]{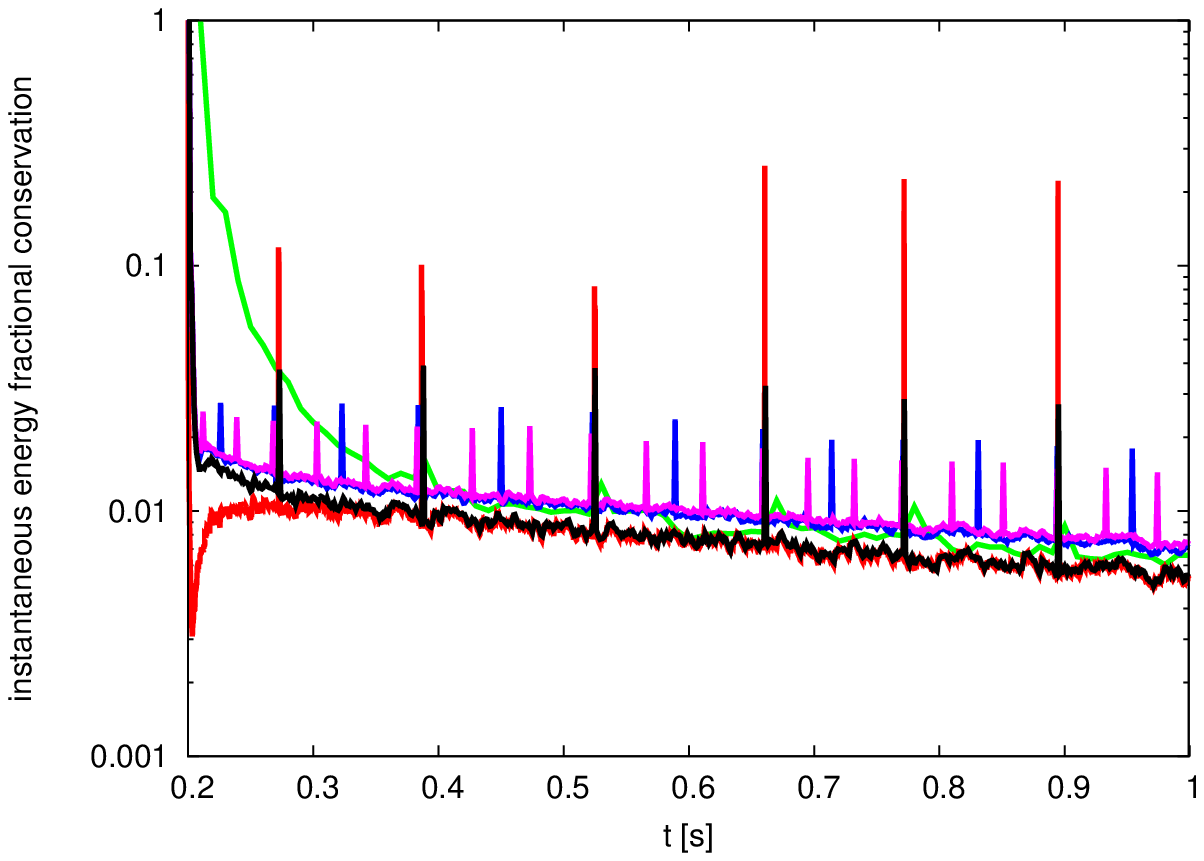}
\caption{Total energy
$E_\mathrm{total}$ (normalized with the stellar initial energy) and
instantaneous energy fractional conservation for a PNS with the CBF-EI EoS and
$\unit[1.60]{M_\odot}$ baryon mass. The timestep \texttt{dt} is kept fixed
during the evolution and \texttt{n} is the number of grid points.
The plots begin at $\unit[200]{ms}$ because this is the initial
time of our simulations (see Sec.~\ref{ssec:evolution_implementation}).
}
\label{fig:conservation_bis}
\end{figure}

The total energy and other quantum numbers (i.e., the baryon number) are
conserved in every physical process. Our code enforces the conservation of the
total baryon number $A=M_\mathrm{B}/m_n$, but as it evolves, the PNS loses
energy and lepton number since neutrinos are allowed to escape from the star.
Since the total energy of a star (matter \emph{plus} neutrinos)
in spherical symmetry is given by its gravitational mass $M$, the total energy
of the system (stellar energy \emph{plus} energy of the emitted neutrinos) is given by
\begin{equation}
E_\mathrm{total}=M(t)+\int_{\unit[200]{ms}}^t L_\nu(t)\mathrm dt,
\end{equation}
where $L_\nu$ is defined in Eq.~\eqref{eq:Lnu_Mdot}.
Similarly, for the electron lepton number,
\begin{align}
N_\mathrm{total}={}&N_L(t) +
\int_{\unit[200]{ms}}^t4\pi R^2\mathrm e^{\phi(R)}F_\nu(R)\mathrm dt,\\
N_L ={}&
\int_0^{A} Y_L(a) \mathrm da,
\end{align}
where $N_L$ is the total number of electronic leptons in the star, and $F_\nu$ is
the electron neutrino number flux (we do not account for the other lepton numbers since
we do not include muons and tauons in the EoSs and moreover
$\mu_{\nu_\mu}=\mu_{\nu_\tau}=0$).

Since the conservation of $E_\mathrm{total}$ and
$N_\mathrm{total}$ has not been enforced, they provide a test for our
simulations. From Fig.~\ref{fig:conservation} and from the top plot of Fig.~\ref{fig:conservation_bis},
it is clear that they are conserved better than about $0.03\%$
during the evolution.

In Fig.~\ref{fig:conservation_bis} we show, for different \emph{fixed}
timesteps and for different grid \emph{dimensions}, the total and instantaneous
energy conservation from 0.2 s to 1 s. The instantaneous energy fractional
conservation is defined as
\begin{equation}
\mathrm{i.e.f.c.}=\frac{|\dot M+L_\nu|}{L_\nu}.
\end{equation}
We see that reducing the timestep the energy conservation is improved. The
instantaneous energy fractional conservation as a function of time shows
regular spikes, whose number doubles (triples) if we double (triple) the grid
points, and whose magnitude is approximately inversely proportional to the
timestep.  
We explain these spikes with the non-linearity of the transport equations (\cite{Press+book},
Sec.~19.1). In fact,
the temperature and the neutrino degeneracy appear inside and outside the
gradients in the transport equations [Eqs.~\eqref{eq:Fnu}--\eqref{eq:dsdt}].
As a consequence, the power in the Fourier space is accumulated in the shorter
wavelengths and is finally released in the longer wavelengths of the solution.
This explains why the frequency of the peaks changes with the grid spacing.
The spikes of the instantaneous energy conservation have magnitudes which
increase when the timestep is lowered, since one is dividing over a smaller
timestep an approximately constant energy jump, $[M(t+\mathrm dt)-M(t)]/\mathrm
dt$.
These spikes do not undermine the overall conservation of the energy and lepton
number and the PNS evolution, see Fig.~\ref{fig:conservation} and upper plot of
Fig.~\ref{fig:conservation_bis}.

\subsection{Beta equilibrium}
\label{ssec:beta-equilibrium}

Our code (as in~\cite{Keil+Janka.1995}) assumes beta equilibrium, Eqs.~\eqref{eq:beta1}.  This approximation is valid if
the timescale of the beta equilibrium is shorter than the dynamical timescale. We estimate the beta equilibrium
timescale using Eqs.~(16) and (17a) of~\cite{Burrows+Lattimer.1986},
\begin{align}
t_\mathrm{beta}={}&\frac{1}{D_n},\\
D_n={}&1.86\times 10^{-2} Y_pT^5[S_4(\eta_e)-S_4(\eta_\nu)]\notag\\
&\cdot\frac{1-\mathrm e^{-\Delta/T}}{1-\mathrm e^{-\eta_e+\eta_\nu}} \unit{\frac{neutrinos}{baryon\cdot s}},\\
S_4(y)={}&\frac{y^5}5+2\pi^2\frac{y^3}{3} + 7\pi^4\frac{y}{15},
\end{align}
where $D_n$ is the net rate of production of electron-neutrinos, $\eta=\mu/T$
is the degeneracy parameter, and $\Delta=0$ in the case of beta equilibrium (we
refer the reader to~\cite{Burrows+Lattimer.1986} for more details). Since we
have assumed beta equilibrium, we put $1-\exp(-\Delta/T)\equiv1$ to estimate
the corresponding timescale. This means that the value of the beta equilibrium
timescale is not fully consistent.

Since the PNS structure changes due to how neutrinos transfer energy and lepton number through the
stellar layers, we estimate the dynamical timescale with the formula
\begin{equation}
t_\mathrm{dyn}=R\frac{n_\nu(r)}{F_\nu(r)},
\end{equation}
where $n_\nu(r)$ and $F_\nu(r)$ are the neutrino number density and number
flux, respectively (which depend on the radial coordinate $r$), and $R$ is the
stellar radius (notice that $F_\nu/n_\nu$ has the dimensions of a velocity).

\begin{figure}
\centering
\includegraphics[width=\columnwidth]{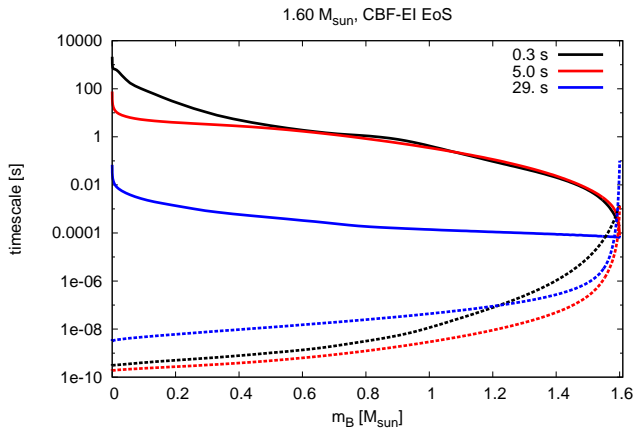}
\caption{Dynamical timescale (solid lines) and beta equilibrium timescale
(dashed lines) profiles at different times for a PNS with the CBF-EI EoS and
$\unit[1.60]{M_\odot}$ stellar mass.}
\label{fig:timescales}
\end{figure}

In Fig.~\ref{fig:timescales} we plot the dynamical and beta equilibrium
timescales for a PNS with the CBF-EI EoS and
$M_\mathrm{B}=\unit[1.60]{M_\odot}$.  The beta equilibrium is valid in most of
the star, apart for a thin shell near the stellar
border.  Towards the end of the evolution the dynamical timescale seems to
reduce, and this is counterintuitive. In fact, we have associated the dynamical
timescale with the neutrino timescale. This is not true towards the end of the
evolution, since as the PNS becomes optically thin the neutrinos decouple from
the matter and the diffusion approximation breaks down.  At that point, the
neutrino timescale drops, but the stellar dynamics is actually frozen.

\subsection{Baryon gas assumption}
\label{ssec:NSE}

\begin{figure}
\centering
\includegraphics[width=\columnwidth]{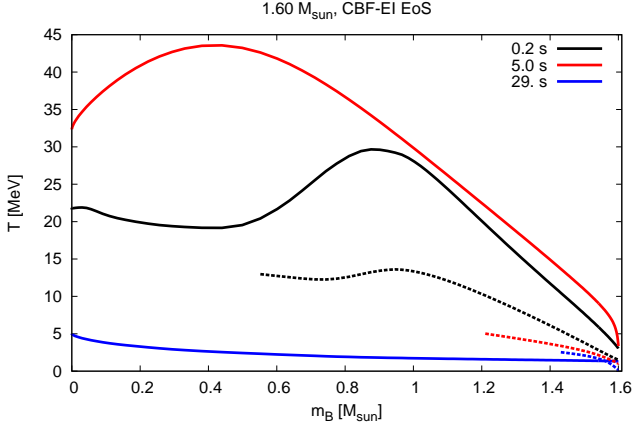}
\caption{Stellar temperature (solid lines) and critical temperature (dashed
lines) for the formation of alpha particles at different times, for a PNS with the CBF-EI EoS and
$\unit[1.60]{M_\odot}$ baryon mass.  When the
baryon density reaches the nuclei density, $\unit[0.155]{fm^{-3}}$, alpha
particles could not form and we do not plot the critical temperature anymore.}
\label{fig:Tc}
\end{figure}

In this paper we have not considered the formation of any kind of crust or
envelope, that is, the EoS baryon part is made by an interacting
gas of protons and neutrons.  However, at low temperature and baryon density,
the matter is not constituted by a gas of baryons only.
The alpha particles (i.e., Helium nuclei) are the first species that appears
decreasing the temperature and the density.  The critical temperature at which
alpha particles begin to form, that is, the
lowest temperature at which protons and neutrons are present alone as an
interacting gas, depends on the baryon density and the proton fraction.
Eq.~(2.31) of~\cite{Lattimer+Swesty.1991} is an estimate of this critical temperature,
\begin{multline}
\label{eq:Tcritical}
T_c(Y_p)=87.76\left(\frac{K_s}{\unit[375]{MeV}}\right)^{1/2}\left(\frac{\unit[0.155]{fm^{-3}}}{n_s}\right)^{1/3}\\
Y_p(1-Y_p)\unit[\,]{MeV},
\end{multline}
where $n_s$ and $K_s$ are the saturation density and the incompressibility
parameter at saturation density of symmetric nuclear matter. Eq.~\eqref{eq:Tcritical} is valid for
$n_\mathrm{B}<n_s$, otherwise no alpha particles may form.  In
Fig.~\ref{fig:Tc} we report the profiles of the critical temperature and the
PNS temperature for different snapshots of a PNS
with the CBF-EI EoS and with $M_\mathrm{B}=\unit[1.60]{M_\odot}$; the results
for the other EoSs and baryon masses are similar. As expected, the assumption of
a proton-neutron interacting gas is valid at the beginning of the
simulation and loses accuracy towards the end of the evolution,
when it is not valid only in the outermost layers.

\section{PNS quasi-normal modes}
\label{app:tables}
In Tables~\ref{tab:GM3_1.25}-\ref{tab:LSbulk_1.60} we report the QNM frequencies and damping
times for the models considered in this paper.

\begin{table}[ht]
\caption{QNMs for a $M_\mathrm{B}=\unit[1.25]{M_\odot}$ star evolved with the GM3 EoS.
The column content, from left to right, is: time of the snapshot (in s), frequency (in Hz) and
damping time (in s)
of the $g_1$-, $f$-, and $p_1$-modes, stellar gravitational mass (in $M_\odot$), and stellar radius (in km).
The $g_1$-mode quantities are not shown when $\tau_{g_1}\gtrsim10^7$ s.}
\label{tab:GM3_1.25}
\centering
\begin{tabular}{ccccccccc}
$t$ & $\nu_{g_1}$ & $\tau_{g_1}$ & $\nu_f$ & $\tau_f$ & $\nu_{p_1}$ & $\tau_{p_1}$ & $M$ & $R$ \\
\hline
0.2 & 784.1 & 4.01 & 955.3 & 2.49 & 1712. & 1.57 & 1.2064 & 22.444 \\
0.3 & 776.6 & 5.91 & 928.2 & 2.18 & 1619. & 1.59 & 1.2051 & 23.530 \\
0.4 & 768.0 & 11.5 & 925.7 & 1.75 & 1623. & 1.56 & 1.2036 & 23.320 \\
0.5 & 752.9 & 25.9 & 940.5 & 1.48 & 1653. & 1.55 & 1.2021 & 22.917 \\
0.6 & 734.4 & 58.2 & 965.2 & 1.31 & 1696. & 1.57 & 1.2006 & 22.430 \\
0.7 & 715.1 & 123. & 995.0 & 1.17 & 1752. & 1.62 & 1.1992 & 21.894 \\
0.8 & 695.8 & 247. & 1027. & 1.05 & 1817. & 1.71 & 1.1977 & 21.349 \\
0.9 & 676.7 & 472. & 1058. & .952 & 1890. & 1.83 & 1.1962 & 20.812 \\
1.0 & 658.2 & 873. & 1090. & .869 & 1976. & 1.98 & 1.1948 & 20.272 \\
2.0 & 504.3 & $1.5\times10^5$ & 1300. & .534 & 3386. & 3.96 & 1.1828 & 15.893 \\
4.0 & 326.8 & $2.8\times10^5$ & 1412. & .446 & 4593. & 5.96 & 1.1697 & 13.502 \\
5.0 & -     & -    & 1433. & .434 & 4756. & 6.46 & 1.1660 & 13.215 \\
10. & -     & -    & 1468. & .415 & 5017. & 6.77 & 1.1562 & 12.778 \\
15. & -     & -    & 1475. & .413 & 5074. & 6.66 & 1.1520 & 12.659 \\
20. & -     & -    & 1473. & .415 & 5091. & 6.88 & 1.1502 & 12.600 \\
\hline
\end{tabular}
\end{table}
\begin{table}[ht]
\caption{As Tab.~\ref{tab:GM3_1.25}, for a $M_\mathrm{B}=\unit[1.25]{M_\odot}$ star with the CBF-EI EoS.}
\label{tab:CBFEI_1.25}
\centering
\begin{tabular}{ccccccccc}
$t$ & $\nu_{g_1}$ & $\tau_{g_1}$ & $\nu_f$ & $\tau_f$ & $\nu_{p_1}$ & $\tau_{p_1}$ & $M$ & $R$ \\
\hline
0.2 & 756.2 & 3.13 & 1036. & 4.41 & 1625. & 1.36 &1.2118 & 23.807 \\
0.3 & 753.2 & 3.35 & 1001. & 4.94 & 1546. & 1.38 &1.2102 & 24.808 \\
0.4 & 767.0 & 3.47 & 976.8 & 4.45 & 1549. & 1.34 &1.2085 & 24.628 \\
0.5 & 783.5 & 3.82 & 959.2 & 3.53 & 1572. & 1.30 &1.2068 & 24.238 \\
0.6 & 796.5 & 4.88 & 950.3 & 2.60 & 1604. & 1.27 &1.2052 & 23.765 \\
0.7 & 800.8 & 7.85 & 952.9 & 1.93 & 1643. & 1.25 &1.2035 & 23.251 \\
0.8 & 795.4 & 15.4 & 967.7 & 1.55 & 1686. & 1.25 &1.2019 & 22.723 \\
0.9 & 782.6 & 32.2 & 990.7 & 1.33 & 1732. & 1.26 &1.2004 & 22.200 \\
1.0 & 766.1 & 66.4 & 1019. & 1.18 & 1783. & 1.28 &1.1988 & 21.679 \\
2.0 & 568.0 & $2.5\times10^4$ & 1296. & .556 & 2542. & 2.32 & 1.1856 & 17.255 \\
4.0 & 377.7 & $3.7\times10^5$ & 1512. & .385 & 4206. & 6.46 & 1.1695 & 13.684 \\
5.0 & - & - & 1553. & .365 & 4643. & 7.75 &1.1647&13.113 \\
10. & - & - & 1635. & .330 & 5355. & 8.57 &1.1535&12.240 \\
15. & - & - & 1659. & .321 & 5507. & 8.04 &1.1499&12.012 \\
20. & - & - & 1664. & .319 & 5568. & 8.00 &1.1492&11.903 \\
\end{tabular}
\end{table}
\begin{table}
\caption{As Tab.~\ref{tab:GM3_1.25}, for a $M_\mathrm{B}=\unit[1.25]{M_\odot}$ star with the LS-bulk EoS.}
\label{tab:LSbulk_1.25}
\centering
\begin{tabular}{ccccccccc}
$t$ & $\nu_{g_1}$ & $\tau_{g_1}$ & $\nu_f$ & $\tau_f$ & $\nu_{p_1}$ & $\tau_{p_1}$ & $M$ & $R$ \\
\hline
0.2 & 870.3 & 8.57 & 1047. & 1.21 & 1982. & 1.23 &1.2020 & 19.925 \\
0.3 & 839.0 & 22.0 & 1045. & 1.12 & 1888. & 1.27 &1.2006 & 20.780 \\
0.4 & 805.6 & 62.2 & 1068. & 1.01 & 1907. & 1.27 &1.1992 & 20.506 \\
0.5 & 772.9 & 161. & 1102. & .911 & 1961. & 1.30 &1.1976 & 20.047 \\
0.6 & 742.2 & 380. & 1139. & .821 & 2035. & 1.37 &1.1961 & 19.533 \\
0.7 & 713.6 & 829. & 1176. & .743 & 2125. & 1.46 &1.1946 & 19.003 \\
0.8 & 686.9 & $1.7\times10^3$ & 1213. & .676 & 2231. & 1.59 &1.1931 & 18.467\\
0.9 & 662.0 & $3.4\times10^3$ & 1248. & .622 & 2353. & 1.74 &1.1917 & 17.948\\
1.0 & 638.7 & $6.5\times10^3$ & 1279. & .579 & 2490. & 1.92 &1.1903 & 17.446\\
2.0 & 473.5 & $5.9\times10^6$ & 1457. & .415 & 4027. & 3.22 &1.1789 & 13.939\\
4.0 & 321.8 & $3.3\times10^5$ & 1538. & .372 & 4864. & 4.89 &1.1673 & 12.557\\
5.0 & 278.9 & $5.9\times10^5$ & 1552. & .366 & 5010. & 5.28 &1.1638 & 12.378\\
10. & - & - & 1572. & .359 & 5267. & 6.01 & 1.1540 & 12.091 \\
15. & - & - & 1575. & .359 & 5328. & 6.15 & 1.1493 & 12.025 \\
20. & - & - & 1575. & .360 & 5352. & 6.22 & 1.1465 & 11.986 \\
\end{tabular}
\end{table}
\begin{table}
\caption{As Tab.~\ref{tab:GM3_1.25}, for a $M_\mathrm{B}=\unit[1.40]{M_\odot}$ star with the GM3 EoS.}
\label{tab:GM3_1.40}
\centering
\begin{tabular}{ccccccccc}
$t$ & $\nu_{g_1}$ & $\tau_{g_1}$ & $\nu_f$ & $\tau_f$ & $\nu_{p_1}$ & $\tau_{p_1}$ & $M$ & $R$ \\
\hline
0.2 & 695.5 &       3.28 & 954.3 & 4.83 & 1495. & 1.36 & 1.3553 & 25.961 \\
0.3 & 712.6 &       3.38 & 924.5 & 4.62 & 1469. & 1.35 & 1.3536 & 26.289 \\
0.4 & 734.6 &       3.67 & 904.4 & 3.61 & 1494. & 1.29 & 1.3518 & 25.782 \\
0.5 & 751.3 &       4.78 & 895.6 & 2.54 & 1530. & 1.25 & 1.3500 & 25.174 \\
0.6 & 757.2 &       8.21 & 901.1 & 1.83 & 1572. & 1.22 & 1.3482 & 24.560 \\
0.7 & 752.5 &       17.2 & 919.9 & 1.46 & 1619. & 1.21 & 1.3464 & 23.937 \\
0.8 & 741.5 &       37.5 & 947.4 & 1.24 & 1672. & 1.22 & 1.3447 & 23.312 \\
0.9 & 727.4 &       78.7 & 979.1 & 1.09 & 1730. & 1.25 & 1.3429 & 22.704 \\
1.0 & 712.1 &       156. & 1012. & .972 & 1793. & 1.29 & 1.3412 & 22.119 \\
2.0 & 567.7 & $3.1\times10^4$ & 1284. & .484 & 2938. & 2.57 & 1.3260 & 17.122 \\
4.0 & 387.1 & $4.6\times10^6$ & 1447. & .372 & 4461. & 4.15 & 1.3084 & 13.827 \\
5.0 & -     &       -    & 1476. & .357 & 4678. & 4.64 & 1.3033 & 13.450 \\
10. & -     &       -    & 1531. & .335 & 5050. & 5.14 & 1.2893 & 12.872 \\
15. & -     &       -    & 1545. & .330 & 5142. & 4.98 & 1.2827 & 12.725 \\
20. & -     &       -    & 1548. & .330 & 5175. & 4.95 & 1.2791 & 12.661 \\
\end{tabular}
\end{table}
\begin{table}
\caption{As Tab.~\ref{tab:GM3_1.25}, for a $M_\mathrm{B}=\unit[1.40]{M_\odot}$ star with the CBF-EI EoS.}
\label{tab:CBFEI_1.40}
\centering
\begin{tabular}{ccccccccc}
$t$ & $\nu_{g_1}$ & $\tau_{g_1}$ & $\nu_f$ & $\tau_f$ & $\nu_{p_1}$ & $\tau_{p_1}$ & $M$ & $R$ \\
\hline
0.2 & 652.7 &            3.67 & 1029. & 6.68 & 1436. & 1.31 & 1.3613 & 27.434 \\
0.3 & 665.1 &            3.60 & 1004. & 7.48 & 1410. & 1.31 & 1.3591 & 27.740 \\
0.4 & 688.6 &            3.39 & 984.5 & 7.13 & 1430. & 1.26 & 1.3570 & 27.266 \\
0.5 & 715.0 &            3.20 & 966.5 & 6.27 & 1462. & 1.21 & 1.3550 & 26.671 \\
0.6 & 741.6 &            3.12 & 950.6 & 5.11 & 1498. & 1.16 & 1.3529 & 26.044 \\
0.7 & 766.0 &            3.25 & 937.7 & 3.84 & 1536. & 1.12 & 1.3510 & 25.421 \\
0.8 & 785.3 &            3.89 & 931.2 & 2.70 & 1578. & 1.09 & 1.3490 & 24.805 \\
0.9 & 795.7 &            5.84 & 934.4 & 1.91 & 1621. & 1.07 & 1.3471 & 24.204 \\
1.0 & 795.4 &            11.0 & 948.7 & 1.48 & 1666. & 1.06 & 1.3453 & 23.622 \\
2.0 & 633.8 & $5.4\times10^3$ & 1239. & .556 & 2308. & 1.50 & 1.3290 & 18.681 \\
4.0 & 381.0 & $3.7\times10^5$ & 1516. & .337 & 4010. & 4.28 & 1.3079 & 14.254 \\
5.0 &  -    &            -    & 1568. & .315 & 4539. & 5.48 & 1.3013 & 13.520 \\
10. &  -    &            -    & 1670. & .280 & 5459. & 7.33 & 1.2844 & 12.420 \\
15. &  -    &            -    & 1703. & .270 & 5674. & 6.92 & 1.2778 & 12.146 \\
20. &  -    &            -    & 1717. & .266 & 5761. & 6.69 & 1.2752 & 12.018 \\
\end{tabular}
\end{table}
\begin{table}
\caption{As Tab.~\ref{tab:GM3_1.25}, for a $M_\mathrm{B}=\unit[1.40]{M_\odot}$ star with the LS-bulk EoS.}
\label{tab:LSbulk_1.40}
\centering
\begin{tabular}{ccccccccc}
$t$ & $\nu_{g_1}$ & $\tau_{g_1}$ & $\nu_f$ & $\tau_f$ & $\nu_{p_1}$ & $\tau_{p_1}$ & $M$ & $R$ \\
\hline
0.2 & 821.7 &            2.70 & 1017. & 2.61 & 1739. & 1.00 & 1.3505 & 22.909 \\
0.3 & 832.1 &            3.58 & 990.7 & 2.07 & 1708. & .993 & 1.3488 & 23.135 \\
0.4 & 834.6 &            6.95 & 988.7 & 1.46 & 1739. & .958 & 1.3470 & 22.658 \\
0.5 & 821.9 &            18.2 & 1010. & 1.16 & 1786. & .939 & 1.3451 & 22.081 \\
0.6 & 800.7 &            49.2 & 1045. & .986 & 1846. & .938 & 1.3433 & 21.461 \\
0.7 & 776.9 &            122. & 1086. & .862 & 1916. & .957 & 1.3415 & 20.833 \\
0.8 & 752.7 &            278. & 1128. & .761 & 1996. & .996 & 1.3396 & 20.219 \\
0.9 & 729.1 &            595. & 1171. & .677 & 2087. & 1.05 & 1.3379 & 19.615 \\
1.0 & 706.3 & $1.2\times10^3$ & 1213. & .608 & 2190. & 1.13 & 1.3361 & 19.035 \\
2.0 & 533.6 & $2.5\times10^5$ & 1472. & .359 & 3688. & 2.10 & 1.3214 & 14.668 \\
4.0 & -     &            -    & 1590. & .305 & 4757. & 3.42 & 1.3058 & 12.757 \\
5.0 & -     &            -    & 1610. & .298 & 4946. & 3.76 & 1.3012 & 12.532 \\
10. & -     &            -    & 1644. & .288 & 5299. & 4.41 & 1.2877 & 12.165 \\
15. & -     &            -    & 1650. & .288 & 5394. & 4.58 & 1.2807 & 12.082 \\
20. & -     &            -    & 1650. & .289 & 5431. & 4.64 & 1.2764 & 12.042 \\
\end{tabular}
\end{table}
\begin{table}
\caption{As Tab.~\ref{tab:GM3_1.25}, for a $M_\mathrm{B}=\unit[1.60]{M_\odot}$ star with the GM3 EoS.}
\label{tab:GM3_1.60}
\centering
\begin{tabular}{ccccccccc}
$t$ & $\nu_{g_1}$ & $\tau_{g_1}$ & $\nu_f$ & $\tau_f$ & $\nu_{p_1}$ & $\tau_{p_1}$ & $M$ & $R$ \\
\hline
0.2 & 548.5 &            4.72 & 946.8 & 10.6 & 1232. & 1.41 & 1.5571 & 32.104 \\
0.3 & 587.0 &            4.05 & 930.4 & 9.24 & 1276. & 1.32 & 1.5546 & 30.898 \\
0.4 & 626.4 &            3.53 & 915.4 & 7.84 & 1329. & 1.22 & 1.5522 & 29.722 \\
0.5 & 664.1 &            3.18 & 901.1 & 6.38 & 1381. & 1.14 & 1.5498 & 28.653 \\
0.6 & 699.0 &            3.04 & 889.0 & 4.82 & 1433. & 1.07 & 1.5475 & 27.675 \\
0.7 & 728.6 &            3.24 & 881.2 & 3.35 & 1484. & 1.02 & 1.5453 & 26.797 \\
0.8 & 749.6 &            4.27 & 881.7 & 2.22 & 1535. & .983 & 1.5431 & 25.993 \\
0.9 & 758.7 &            7.45 & 894.3 & 1.57 & 1586. & .958 & 1.5410 & 25.236 \\
1.0 & 757.3 &            15.6 & 917.9 & 1.25 & 1640. & .946 & 1.5388 & 24.525 \\
2.0 & 636.9 & $5.4\times10^3$ & 1239. & .464 & 2481. & 1.46 & 1.5195 & 18.838 \\
4.0 & 458.1 & $6.7\times10^5$ & 1484. & .305 & 4246. & 2.61 & 1.4949 & 14.259 \\
5.0 & 397.8 & $10^6$          & 1526. & .289 & 4531. & 3.03 & 1.4877 & 13.735 \\
10. &  -    &            -    & 1611. & .263 & 5050. & 3.69 & 1.4675 & 12.955 \\
15. &  -    &            -    & 1636. & .256 & 5210. & 3.62 & 1.4572 & 12.750 \\
20. &  -    &            -    & 1646. & .254 & 5261. & 3.51 & 1.4509 & 12.665 \\
\end{tabular}
\end{table}
\begin{table}
\caption{As Tab.~\ref{tab:GM3_1.25}, for a $M_\mathrm{B}=\unit[1.60]{M_\odot}$ star with the CBF-EI EoS.}
\label{tab:CBFEI_1.60}
\centering
\begin{tabular}{ccccccccc}
$t$ & $\nu_{g_1}$ & $\tau_{g_1}$ & $\nu_f$ & $\tau_f$ & $\nu_{p_1}$ & $\tau_{p_1}$ & $M$ & $R$ \\
\hline
0.2 & 522.7 &            7.45 & 995.1 & 20.4 & 1197. & 1.41 & 1.5637 & 33.618 \\
0.3 & 549.3 &            5.01 & 986.9 & 15.3 & 1231. & 1.36 & 1.5606 & 32.492 \\
0.4 & 581.2 &            4.17 & 978.8 & 12.3 & 1273. & 1.29 & 1.5577 & 31.376 \\
0.5 & 614.4 &            3.62 & 969.0 & 10.4 & 1318. & 1.21 & 1.5551 & 30.341 \\
0.6 & 647.5 &            3.21 & 957.7 & 8.95 & 1362. & 1.14 & 1.5525 & 29.389 \\
0.7 & 680.3 &            2.90 & 945.4 & 7.56 & 1408. & 1.08 & 1.5500 & 28.496 \\
0.8 & 711.9 &            2.69 & 933.3 & 6.14 & 1453. & 1.02 & 1.5476 & 27.677 \\
0.9 & 741.4 &            2.62 & 922.2 & 4.67 & 1497. & .978 & 1.5452 & 26.910 \\
1.0 & 767.2 &            2.80 & 914.3 & 3.28 & 1542. & .941 & 1.5430 & 26.187 \\
1.1 & 786.1 &            3.58 & 913.0 & 2.20 & 1588. & .911 & 1.5407 & 25.510 \\
1.2 & 794.6 &            5.98 & 921.9 & 1.55 & 1634. & .889 & 1.5386 & 24.860 \\
1.3 & 792.6 &            12.2 & 941.4 & 1.22 & 1682. & .874 & 1.5364 & 24.234 \\
1.4 & 783.8 &            26.1 & 967.9 & 1.04 & 1732. & .867 & 1.5343 & 23.633 \\
1.8 & 725.5 &            365. & 1095. & .691 & 1957. & .910 & 1.5263 & 21.458 \\
1.9 & 709.4 &            648. & 1127. & .635 & 2022. & .939 & 1.5244 & 20.962 \\
2.0 & 693.2 & $1.1\times10^3$ & 1158. & .586 & 2090. & .975 & 1.5226 & 20.486 \\
4.0 & 453.8 & $5.5\times10^5$ & 1522. & .291 & 3758. & 2.55 & 1.4944 & 14.944 \\
5.0 & -     &               - & 1590. & .267 & 4348. & 3.37 & 1.4853 & 14.006 \\
10. & -     &               - & 1714. & .233 & 5492. & 5.69 & 1.4600 & 12.639 \\
15. & -     &               - & 1755. & .224 & 5801. & 5.88 & 1.4485 & 12.295 \\
20. & -     &               - & 1776. & .220 & 5926. & 5.64 & 1.4425 & 12.143 \\
\end{tabular}
\end{table}
\begin{table}
\caption{As Tab.~\ref{tab:GM3_1.25}, for a $M_\mathrm{B}=\unit[1.60]{M_\odot}$ star with the LS-bulk EoS.}
\label{tab:LSbulk_1.60}
\centering
\begin{tabular}{ccccccccc}
$t$ & $\nu_{g_1}$ & $\tau_{g_1}$ & $\nu_f$ & $\tau_f$ & $\nu_{p_1}$ & $\tau_{p_1}$ & $M$ & $R$ \\
\hline
0.2 & 662.9 &            3.07 & 1020. & 6.83 & 1440. & .997 & 1.5519 & 28.240 \\
0.3 & 709.0 &            2.72 & 997.3 & 5.89 & 1494. & .919 & 1.5495 & 27.040 \\
0.4 & 753.6 &            2.52 & 976.8 & 4.56 & 1555. & .846 & 1.5471 & 26.001 \\
0.5 & 792.9 &            2.64 & 962.0 & 3.06 & 1618. & .787 & 1.5448 & 25.029 \\
0.6 & 819.9 &            3.71 & 959.7 & 1.86 & 1680. & .745 & 1.5424 & 24.138 \\
0.7 & 827.0 &            8.06 & 977.4 & 1.25 & 1743. & .717 & 1.5401 & 23.325 \\
0.8 & 818.3 &            21.6 & 1012. & .988 & 1809. & .701 & 1.5379 & 22.547 \\
0.9 & 802.1 &            55.8 & 1055. & .838 & 1879. & .699 & 1.5356 & 21.818 \\
1.0 & 782.9 &            132. & 1102. & .728 & 1956. & .708 & 1.5334 & 21.124 \\
2.0 & 606.4 & $5.3\times10^4$ & 1473. & .315 & 3237. & 1.24 & 1.5140 & 15.752 \\
4.0 & 426.3 & $1.3\times10^5$ & 1657. & .245 & 4586. & 2.17 & 1.4924 & 12.975 \\
5.0 & 374.9 & $9.2\times10^6$ & 1687. & .237 & 4825. & 2.44 & 1.4860 & 12.678 \\
10. &  -    &            -    & 1742. & .225 & 5304. & 3.03 & 1.4670 & 12.201 \\
15. &  -    &            -    & 1755. & .223 & 5452. & 3.19 & 1.4568 & 12.081 \\
20. &  -    &            -    & 1757. & .224 & 5510. & 3.25 & 1.4500 & 12.032 \\
\end{tabular}
\end{table}

\bibliography{bibliografia}

\end{document}